\documentclass[iop]{emulateapj}

\usepackage{appendix,natbib}
\usepackage{amsmath}
\usepackage{subfigure}
\usepackage[backref,colorlinks=true,citecolor=blue,linkcolor=blue,urlcolor=blue]{hyperref}
\usepackage[all]{hypcap}

\usepackage{amsmath}
\usepackage{appendix}
\usepackage[utf8]{inputenc}
\usepackage{listingsutf8}
\usepackage{subfigure}

\newcommand{\halpha}{H\ensuremath{\alpha}}
\newcommand{\hbeta}{H\ensuremath{\beta}}
\newcommand{\um}{\,$\mu$m}
\def\o32{{O$_{\rm 32}$}}
\def\msun{\,${M_\odot}$}
\def\lsun{\,${\rm L_\odot}$}
\def\sfr{{\rm SFR$_{\rm{H\alpha,H\beta}}$}}
\def\taub{$\tau_{\rm b}$}
\slugcomment{Draft: {\today} -- accepted for publication in ApJ}

\begin{document}

\title{The {MOSDEF} Survey: the Variation of the Dust Attenuation Curve with Metallicity}

\author{\sc Irene Shivaei\altaffilmark{1,2}, Naveen Reddy\altaffilmark{3}, George Rieke\altaffilmark{1}, Alice Shapley\altaffilmark{4}, Mariska Kriek\altaffilmark{5}, Andrew Battisti\altaffilmark{6}, Bahram Mobasher\altaffilmark{3}, Ryan Sanders\altaffilmark{7}, Tara Fetherolf\altaffilmark{3}, Mojegan Azadi\altaffilmark{8}, Alison L. Coil\altaffilmark{9}, William R. Freeman\altaffilmark{3}, Laura de Groot\altaffilmark{10}, Gene Leung\altaffilmark{9}, Sedona H. Price\altaffilmark{11}, Brian Siana\altaffilmark{3}, and Tom Zick\altaffilmark{4}}
\altaffiltext{1}{Steward Observatory, University of Arizona, Tucson, AZ 85721, USA}
\altaffiltext{2}{NASA Hubble fellow}
\altaffiltext{3}{Department of Physics \& Astronomy, University of California, Riverside, CA 92521, USA}
\altaffiltext{4}{Department of Physics \& Astronomy, University of California, Los Angeles, CA 90095, USA}
\altaffiltext{5}{Astronomy Department, University of California, Berkeley, CA 94720, USA}
\altaffiltext{6}{Research School of Astronomy and Astrophysics, Australian National University, Cotter Road, Weston Creek, ACT 2611, Australia}
\altaffiltext{7}{Department of Physics, University of California, Davis, One Shields Ave, Davis, CA 95616, USA}
\altaffiltext{8}{Harvard-Smithsonian Center for Astrophysics, 60 Garden Street, Cambridge, MA 02138, USA}
\altaffiltext{9}{Center for Astrophysics and Space Sciences, University of California, San Diego, La Jolla, CA 92093, USA}
\altaffiltext{10}{Department of Physics, The College of Wooster, 1189 Beall Avenue, Wooster, OH 44691, USA}
\altaffiltext{11}{Max-Planck-Institut f\"{u}r extraterrestrische Physik, Postfach 1312, Garching, 85741, Germany}

\begin{abstract}

We derive the UV-optical stellar dust attenuation curve of galaxies at $z=1.4-2.6$ as a function of gas-phase metallicity. We use a sample of 218 star-forming galaxies, excluding those with very young or heavily obscured star formation, from the MOSFIRE Deep Evolution Field (MOSDEF) survey with {\halpha}, {\hbeta}, and [N{\sc ii}]$\lambda 6585$ spectroscopic measurements. We constrain the shape of the attenuation curve by comparing the average flux densities of galaxies sorted into bins of dust obscuration using Balmer decrements, i.e., {\halpha}-to-{\hbeta} luminosities. 
The average attenuation curve for the high-metallicity sample ($12+\log({\rm O/H})>8.5$, corresponding to $M_*\gtrsim10^{10.4}$\,{\msun}) has a shallow slope, identical to that of the Calzetti local starburst curve, and a significant UV 2175\,{\AA} extinction bump that is $\sim 0.5\times$ the strength of the Milky Way bump. On the other hand, the average attenuation curve of the low-metallicity sample ($12+\log({\rm O/H})\sim 8.2-8.5$) has a steeper slope similar to that of the SMC curve, only consistent with the Calzetti slope at the $3\sigma$ level. The UV bump is not detected in the low-metallicity curve, indicating the relative lack of the small dust grains causing the bump at low metallicities. Furthermore, we find that on average the nebular reddening ($E(B-V)$) is a factor of 2 times larger than that of the stellar continuum for galaxies with low metallicities, while the nebular and stellar reddening are similar for galaxies with higher metallicities. The latter is likely due to a high surface density of dusty clouds embedding the star forming regions but also reddening the continuum in the high-metallicity galaxies.

\end{abstract}

\maketitle

\section{Introduction}
What kind of dust attenuation curve should be used to correct the stellar continuum emission of galaxies?
This question has a long history in galaxy evolution studies. 
In spite of the profound differences observed in the extinction curves of the Milky Way \citep[MW;][]{cardelli89,fitzpatrick99} and a few well-studied local galaxies \citep[e.g.,][]{gordon03,fm07,clayton15}, at high redshifts a single attenuation curve is typically adopted to correct the rest-frame UV-optical light.
The most common attenuation curves used at high redshift are the empirical starburst curve of \citet[][hereafter the C00 curve]{calzetti00} and the two-component dust model of \citet{charlotfall00}. However, a variety of other curves are also found in the literature, as discussed below.

Extinction and attenuation curves are different by definition. An extinction curve is measured by observing stars with a screen of dust along the line of sight, which is influenced by the properties of dust grains (i.e., composition, size distribution). An attenuation curve refers to the average impact of dust absorption and scattering on a collection of stars, which includes the effects of the geometrical distribution of dust and stars in addition to the dust grain properties (i.e., as for the extinction). The main differences observed in the shapes of the attenuation/extinction curves are the strengths of a broad extinction feature at 2175\,\AA, called the UV extinction bump, and the steepness (slope) of the UV curve \citep[see the review by][]{salim20}. The bump strength and the UV slope can be affected by both the radiative transfer effects (including dust-star geometry) in the attenuation curve, and by the underlying dust grain properties of the extinction curve.

A variety of behavior is observed in the attenuation/extinction curves in the local universe and over cosmic time. The UV bump is observed along many sightlines within the MW, and to the LMC \citep{fm07,gordon03}, and M31 \citep{bianchi96}, but is weak or absent in the SMC. The bump is also absent in the C00 and the \citet{charlotfall00} attenuation curves.
The SMC extinction curve has a steep UV rise, while the C00 attenuation curve and the MW extinction curve have shallower slopes. 

There are several studies of the attenuation curve of local galaxies that reach a variety of conclusions.
In a series of papers, Battisti and collaborators examined variations in the shape of the attenuation curve for a large sample of local ($z<0.1$) galaxies, using the methodology of \citet{calzetti94}. Accordingly, these authors compared the average spectral energy distributions (SEDs) of galaxies with similar stellar populations but different dust reddening inferred from Balmer decrements, based on the flux ratio of {\halpha} to {\hbeta} \citep{battisti16,battisti17a,battisti17b}. They did not find any trend between the slope of the curve and stellar mass, specific star-formation rate (sSFR), or metallicity. However, they found that galaxies with the highest inclination have a weak UV extinction bump and concluded that overall the effective attenuation ($k_{\rm \lambda}-R_{\rm V}$) in the UV is lower than that derived from the C00 curve.
\citet{wild11} compared the SEDs of pairs of galaxies matched to have similar intrinsic SEDs but different amounts of dust. They found that the UV slope of the curve and the UV bump strength vary strongly with inclination, but not with sSFR or optical depth. They observed the UV bump in all of their samples (except possibly the high sSFR sample). 
In another study at low redshifts ($z=0.01-0.05$), \citet{conroy-schiminovich10} speculated that the UV bump may be responsible for their observed trends in the UV and optical colors. The bump has been also detected in the nearby sample of \citet{burgarella05}, who suggested an average attenuation curve with the characteristics of the LMC curve. 
Using energy-balance SED fitting codes and UV to IR data for samples of local galaxies, \citet{leja17} and \citet{salim18} found a range of attenuation curve slopes, and a correlation between the slope and optical attenuation ($A_V$). \citet{salim18} reported slopes that are on average steep (SMC-like) and a variety of UV bump amplitudes, with an average of $\sim 1/3$ the strength of the MW bump. They also found that the local analogs of high-redshift galaxies have UV attenuation curves even steeper than the SMC curve.

At higher redshifts, $z\sim 1-3$, different methods are used to determine the shape of the attenuation curve, again with a range of results.
\citet{noll09} set direct observational constraints on the strength of the bump by using high-quality rest-frame UV spectra at $z=1.0-2.5$. They concluded that a significant bump is present in at least 30\% of their sample, particularly those with higher reddening. 
\citet{reddy15} used Balmer decrements to empirically derive the attenuation curve at $z\sim 1.5-2.5$. They found marginal evidence for the UV bump and a UV curve slope identical to that of the C00 curve, but with a lower normalization ($R_V$).
In multiple studies, flexible SED fitting techniques that allow the attenuation curve to be a free parameter have been used to infer the shape of the curve. For a sample of $z\sim1.5$ galaxies with high average optical attenuation ($\langle A_V \rangle \sim 0.9$), \citet{buat11} found a shallow curve with a UV bump of $\sim 0.5$ the strength of the MW bump. In a later study, \citet{buat12} used a UV-selected sample with $\langle A_V \rangle \sim 0.5$ and found an average attenuation slope steeper than that of the C00 curve slope and a UV bump with amplitude similar to the LMC super-shell.
Investigating the composite SEDs of galaxies at $z=0.5-2.0$ with $A_V \gtrsim 0.5$, \citet{kriek13} derived an average curve with a shallow slope similar to the C00 slope. They also found that steeper curves show stronger UV bumps, and that the bump strength and curve slope vary with sSFR. 
 \citet{zeimann15} defined an ``intrinsic'' SED using the average properties of their sample, and also found a varying UV slope with stellar mass, and no UV bump. 
\citet{salmon16} used a Bayesian formalism and incorporated IR luminosities along with UV-optical photometry, and found a variety of slopes. These authors showed that galaxies with higher color excesses ($E(B-V)$) have shallower curves.
\citet{tress18} found a wide range of UV bump strengths at $z=1.5-3$ with a correlation between the bump amplitude and color excess.
\citet{scoville15} also observed a weak UV bump and a C00-like curve slope in their sample of young galaxies with $\langle A_V \rangle \sim 0.6$ (identified by strong C{\sc iv}$\lambda 1540$ absorption features) at $z=2.0-6.5$. 
\citet{battisti20} showed that a UV bump of $\sim 1/3$ the strength of the MW bump is required to achieve better SED fits to galaxies at $z\sim 0.1-3.0$, and that the slope of the curve strongly anti-correlates with star-formation rate (SFR) and $A_V$.

The UV extinction bump has also been detected in the afterglow spectrum of Gamma-ray bursts (GRBs) at $z\sim 2$ \citep{eliasdottir09,perley11,zafar18a}. In another study of seven GRB hosts, \citet{corre18} found a great variety in the slopes and UV bump strengths. Multiple other GRB afterglow studies at $z\sim 1-3$ and higher ($z>3$), found steep and bump-free extinction curves consistent with the SMC curve \citep{japelj15,zafar18c}.

In summary, the attenuation curves reported in the literature show a large degree of variation. 
However, in most cases, the variations can be attributed to sample selection effects -- specifically to differences in the average optical attenuation, $A_V$, of the samples \citep{salim20}. Theoretical work based on radiative transfer modeling has shown that the attenuation curve slope and the bump strength correlate with total dust column density, such that galaxies with larger optical depths (or $A_V$) have shallower slopes and suppressed bumps \citep{wittgordon00,pierni04,inoue05,chevallard13}. These relationships are also recovered in cosmological simulations \citep{narayanan18,trayford20,shen20}.

In addition, the attenuation curve is also affected by dust grain properties. The origin of the UV bump is most commonly attributed to small carbonaceous grains in the form of graphite or polycyclic aromatic hydrocarbon \citep[PAHs;][]{stecher65,draine03,weingartner01a,li01}, and the steep UV rise of the extinction curve is speculated to be a result of a higher fraction of small dust grains \citep{gordon97,hou17} and/or a lower fraction of small carbonaceous to small silicate grains \citep{weingartner01a}. Therefore, the bump and the steep UV slope will be more prominent in environments that are amenable to the carriers of the bump and the steep UV rise, respectively. A proxy to probe the small dust grains in the ISM is gas-phase metallicity, owing to dust depletion of interstellar elements \citep{dwek98,jenkins09}. 

In this study, we aim to investigate the variation of the attenuation curve with gas-phase metallicity by fixing the distribution of dust optical depths in our samples to control for the radiative transfer effects and eliminate the selection biases due to different optical depth distributions. 
We adopt the empirical Balmer decrement method that was first developed by \citet{kinney94} and \citet{calzetti94,calzetti00}, and later exploited by \citet{battisti16,battisti17a,battisti17b} for local samples of galaxies and by \citet{reddy15} for high-redshift galaxies. In this method, the Balmer decrement is used as an independent indicator of dust reddening to sort galaxies into bins of obscuration, allowing comparison of the average wavelength-dependent flux densities of more obscured galaxies to those of galaxies that are less obscured. 
Such a method requires a large spectroscopic sample with measurements of at least two Balmer lines, which is achievable at high redshifts using multi-object near-IR spectrometers. 

The MOSFIRE Deep Evolution Field (MOSDEF) survey \citep{kriek15} is one of the largest rest-frame optical spectroscopic surveys of galaxies at $z\sim 1.5-3.5$, with Keck/MOSFIRE spectra of $\sim 1500$ galaxies, covering {\halpha} and {\hbeta} for those at $z\sim 1.5-2.5$. Using the first two (out of four) years of the MOSDEF survey dataset, \citet{reddy15} derived the $z\sim 2$ attenuation curve adopting the Balmer decrement method.
In this study, we build on this previous work by dividing the full MOSDEF sample into two metallicity bins with matched distributions in Balmer optical depth ($\tau_{\rm H\alpha}-\tau_{\rm H\beta}$) and sSFR. The former avoids biases due to different dust column densities, and the latter ensures similar intrinsic stellar populations among the metallicity bins.
Moreover, this study has the advantage of better sampling the rest-frame UV SED compared to similar photometric studies at $z\sim 0$ (usually limited to the {\em GALEX} NUV and FUV bands at $1771-2831$\,{\AA} and $1344-1786$\,{\AA}, respectively), as the rest-frame UV at $z\sim 2$ is redshifted to wavelengths accessible to optical instruments.

The paper is structured as follows. Section~\ref{sec:method} presents the methodology for deriving the attenuation curve. In Section~\ref{sec:data}, we describe the data, sample, and measurements. We present the derivation of the attenuation curve in bins of metallicity in Section~\ref{sec:curves}. In Section~\ref{sec:implications}, we discuss the physical interpretation of our metallicity-dependent attenuation curves, and review the implications of using these curves for calculating dust-corrected SFRs, UV continuum slopes, and stellar color excesses ($E(B-V)$).
Section~\ref{sec:summary} summarizes the main results. 
Throughout this paper, line wavelengths are presented in vacuum. We assume a \citet{chabrier03} IMF and adopt a cosmology with $H_0 = 70\,{\rm km\,s^{-1}\,Mpc^{-1}, \Omega_{\Lambda} = 0.7}$, and ${\rm \Omega_m = 0.3}$.

\section{Methodology}
\label{sec:method}
We follow the methodology that was used initially in \citet{calzetti94}, and later in \citet{calzetti00}, \citet{reddy15}, and \citet{battisti16}, to derive the selective attenuation curve, $k_{\lambda}-R_V$, where $k_{\lambda}$ is the total attenuation curve as a function of wavelength and $R_V$ is the normalization. The normalization is conventionally taken at 5500\,{\AA}, the effective wavelength of the $V$ band (i.e., $k_V=R_V$). Later in Section~\ref{sec:total}, we combine our results on the selective attenuation curve with those of \citet{reddy15} to obtain the total attenuation curve, $k_{\lambda}$. Total attenuation at a given wavelength is then $A_{\lambda} = E(B-V)\,k_{\lambda}$, where $E(B-V)$ is the stellar color excess, or reddening. Assuming a geometry of a screen of dust in front of the emitters, the intrinsic stellar emission ($F_{\lambda,{\rm int}}$) can be recovered using the standard formulation of $F_{\lambda,{\rm int}}=F_{\lambda,{\rm obs}}\times 10^{0.4\,E(B-V)\,k_{\lambda}}$ \citep{calzetti94,calzetti00}, where $F_{\lambda,{\rm obs}}$ is the observed flux density.

The difference in optical depth between {\halpha} and {\hbeta} is the Balmer optical depth, {\taub} \citep{calzetti94}, defined as 
\begin{equation} \label{eq:taub}
\tau_{\rm b}=\tau_{\rm H\alpha}-\tau_{\rm H\beta}= \ln(\frac{{\rm F(H\alpha)/F(H\beta)}}{2.86}),
\end{equation} 
where F({\halpha}) and F({\hbeta}) are the {\halpha} and {\hbeta} line fluxes and 2.86 is the theoretical value for the ratio in Case B recombination with $T_e=10^4$\,K and $n_e=100\,{\rm cm^{-3}}$ \citep{osterbrock06}. 
The Balmer optical depth indicates the nebular dust reddening, which correlates with stellar dust reddening, as shown in multiple previous studies \citep[e.g.,][]{calzetti94,reddy15,battisti16,theios19}. We show the correlation in our sample later in Section~\ref{sec:metal-sample} and Figure~\ref{fig:taub-beta}. 

We sort the galaxies into bins of {\taub}. The lowest {\taub} bin is consistent with $\langle${\taub}$\rangle\sim 0$, and its average flux densities are taken as the ``intrinsic'' (dust-free) SED of the sample. 
The flux densities of each target are normalized at rest-frame 5500\,\AA, using the individual best-fit SEDs (Section~\ref{sec:sed-sfr}), and are shifted to the rest frame. We use the moving median (as a smoothing technique) of flux densities in each {\taub} bin at $\lambda= 1250-7000$\,{\AA} to construct empirical SEDs.
By dividing the empirical SED in each {\taub} bin by that of the lowest {\taub} bin ($\langle${\taub}$\rangle\sim 0$), and normalizing to the difference in Balmer optical depths, we calculate the selective attenuation curve as
\begin{equation}
\label{eq:qi}
Q_i=-~\frac{\ln(\frac{F_{\lambda,i}}{F_{\lambda,0}})}{\tau_{{\rm b},i}-\tau_{\rm b,0}},
\end{equation}
where $\tau_{b,0}$ and $F_{\lambda,0}$ are the average Balmer optical depth and flux density of the lowest {\taub} bin, respectively ($\tau_{b,0}$ is consistent with 0), and the quantities with $i$ indices denote those of the $i$-th {\taub} bins. We adopt three {\taub} bins (two $Q_i$s), as our sample size does not allow for higher number of bins without each bin getting dominated by the statistical noise (read about the sample in Section~\ref{sec:sample}). The average selective curve, or the ``effective'' attenuation ($Q_{\rm eff}$) is the average of all the selective attenuation curves ($Q_i$s). 

To derive robust uncertainties we performed a Monte Carlo simulation by a) randomly perturbing the flux densities assuming a Gaussian distribution centered at the measured value with a standard deviation equal to the photometric error, b) randomly perturbing the boundaries of the {\taub} bins assuming Gaussian distributions centered at the boundaries shown in the right panel of Figure~\ref{fig:sample-distributions}, with a standard deviation of 0.1, which is about the typical error in {\taub} measurements, and c) performing a jackknife resampling \citep{quenouille49,tukey58} by randomly removing an object from the sample without a replacement. The realizations are measured at least 1000 times. The final effective curve and its uncertainty are determined based on the average and standard deviation of the 1000 realizations.

To find the functional form of the attenuation curve, we fit $Q_{\rm eff}$ versus $1/\lambda$ with a second-order polynomial, and, where applicable, a Drude function is added to account for the UV extinction bump using the \citet{fm07} formalization:
\begin{eqnarray}
\label{eq:qeff}
Q_{\rm eff}(x)=a_1 x^2+a_2 x+a_3+D(x,E_b),\\
D(x,E_b)=E_b \frac{x^2}{(x^2-x_0^2)^2+x^2 \gamma^2},
\label{eq:drude}
\end{eqnarray}
where $x=1/\lambda$ in {\um}$^{-1}$ and $D(x,E_b)$ is the Drude function. We fix the central wavelength ($x_0$) and width ($\gamma$) of the UV extinction bump to the MW values of 4.59{\um}$^{-1}$ and 0.922{\um}$^{-1}$, adopted from \citet{fm07}, and leave the amplitude of the Drude function ($E_b$) as a free parameter. The fits are performed at $\lambda=1250-3000$\,{\AA} and $4000-7000$\,{\AA}. Wavelengths of $\lambda<1250$\,{\AA} and $\lambda=3000-4000$\,{\AA} are excluded to avoid variations in Ly$\alpha$ equivalent width and the strength of the Balmer/4000\,{\AA} break, respectively. Our second-order polynomial fit is supported over a linear fit according to the Akaike Information Criterion \citep[AIC;][]{akaike73} and the reduced $\chi^2$ of the fits. A linear fit to $Q_{\rm eff}$ versus $1/\lambda$ always results in a curve with a steep UV slope, as the fit is highly influenced by the longer wavelength data ($\gtrsim 4000$\,{\AA}) and does not have the flexibility to simultaneously fit the UV data points.

The $Q_{\rm eff}$ curve can be related to the total attenuation through 
\begin{equation}
k_{\lambda}\equiv f\,Q_{\rm eff}(\lambda)+R_V.
\end{equation}
Calculating $R_V$, which in this formalization is the vertical offset at 5500\,\AA, requires extrapolation to rest-frame near-IR wavelengths, where the attenuation is considered to be negligible  \citep{calzetti97,gordon03,reddy15,battisti17a}, or alternatively, requires an energy balance calculation between the IR dust emission and UV-optical attenuation \citep{calzetti00}. Constraining the $R_V$ for our curves is beyond the scope of this work and will be discussed in a future paper. Here, we focus on the shape of the attenuation curve at rest-frame UV and optical wavelengths. 
The factor $f$ in the equation above constrains the slope of the curve at $\lambda=4400-5500\,$\AA~and is conventionally defined such that $k_B-k_V=1$:
\begin{equation}
\label{eq:f}
f=\frac{1}{Q_{\rm eff}(B)-Q_{\rm eff}(V)},
\end{equation} 
where the $B$ and $V$ bands correspond to $\lambda=4400$\,{\AA} and 5500\,{\AA}, respectively. Note that in the formalization presented here, by definition, $Q_{\rm eff}(V)$ is zero.

\section{Data and Sample}
\label{sec:data}
\subsection{The MOSDEF Survey}
MOSDEF is a near-IR spectroscopic survey of $\sim 1500$ galaxies at $z=1.4-3.8$ with the Keck/MOSFIRE spectrograph \citep{mclean10,mclean12}, carried out over 48.5 nights of observation from $2012-2016$. The parent sample is selected in three redshift bins of $z=1.37-1.70$, $2.09-2.61$, and $2.95-3.80$, down to $H$-band AB magnitudes of $H=24.0$, 24.5, and 25.0 respectively, corresponding to $M_*\sim 10^9$\,{\msun} for star-forming galaxies, using the photometric and spectroscopic catalogs from the 3D-{\em HST} survey \citep{skelton14,momcheva16}. The three redshift intervals ensure coverage of the strongest rest-optical nebular emission lines, including [O{\sc ii}], {\hbeta}, [O{\sc iii}], and {\halpha}. As the {\halpha} line is not accessible for the highest redshift bin, only the first two redshift bins ($z=1.37-2.61$) are used in this analysis. The targets for MOSFIRE observations are further selected based on prior redshift measurements (external spectroscopic or 3D-{\em HST} grism redshifts and/or photometric redshifts) that place them in the aforementioned redshift ranges. We refer readers to \citet{kriek15} for more details on the survey strategy, observations, and data reduction.
 
MOSDEF targets lie in the deep extragalactic CANDELS fields: AEGIS, COSMOS, GOODS-N, GOODS-S, and UDS \citep{grogin11,koekemoer11}, and are hence covered by a wealth of ancillary data. In this work, the UV to near-IR photometric catalogs of the 3D-{\em HST} survey \citep{skelton14} are used for SED fitting and deriving rest-frame UV SFRs and UV slopes (Section~\ref{sec:sed-sfr}).

\begin{figure*}[t]
		\includegraphics[width=.33\textwidth,trim={.2cm 0 0 0},clip]{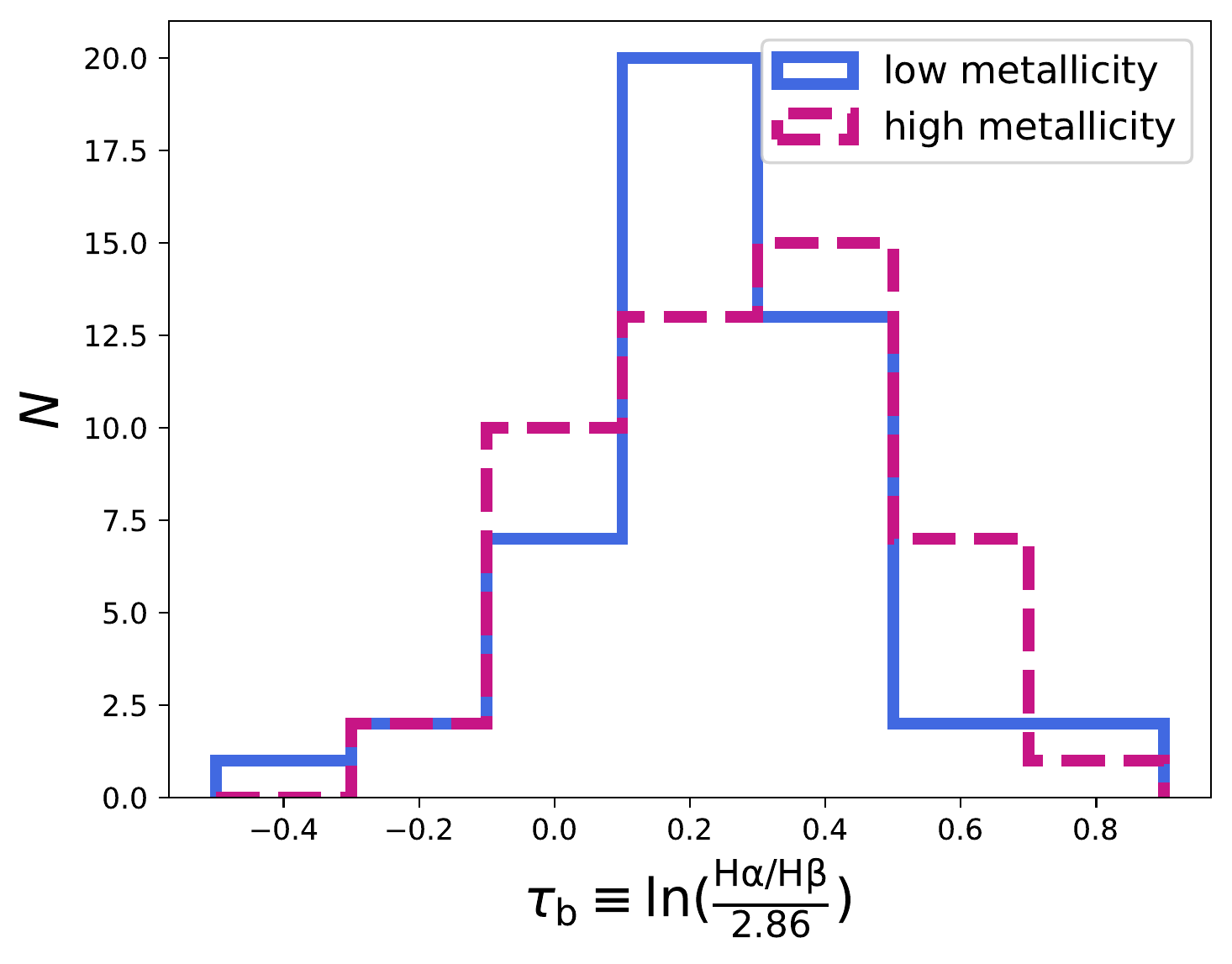} \quad
		\includegraphics[width=.33\textwidth,trim={.2cm 0 0 0},clip]{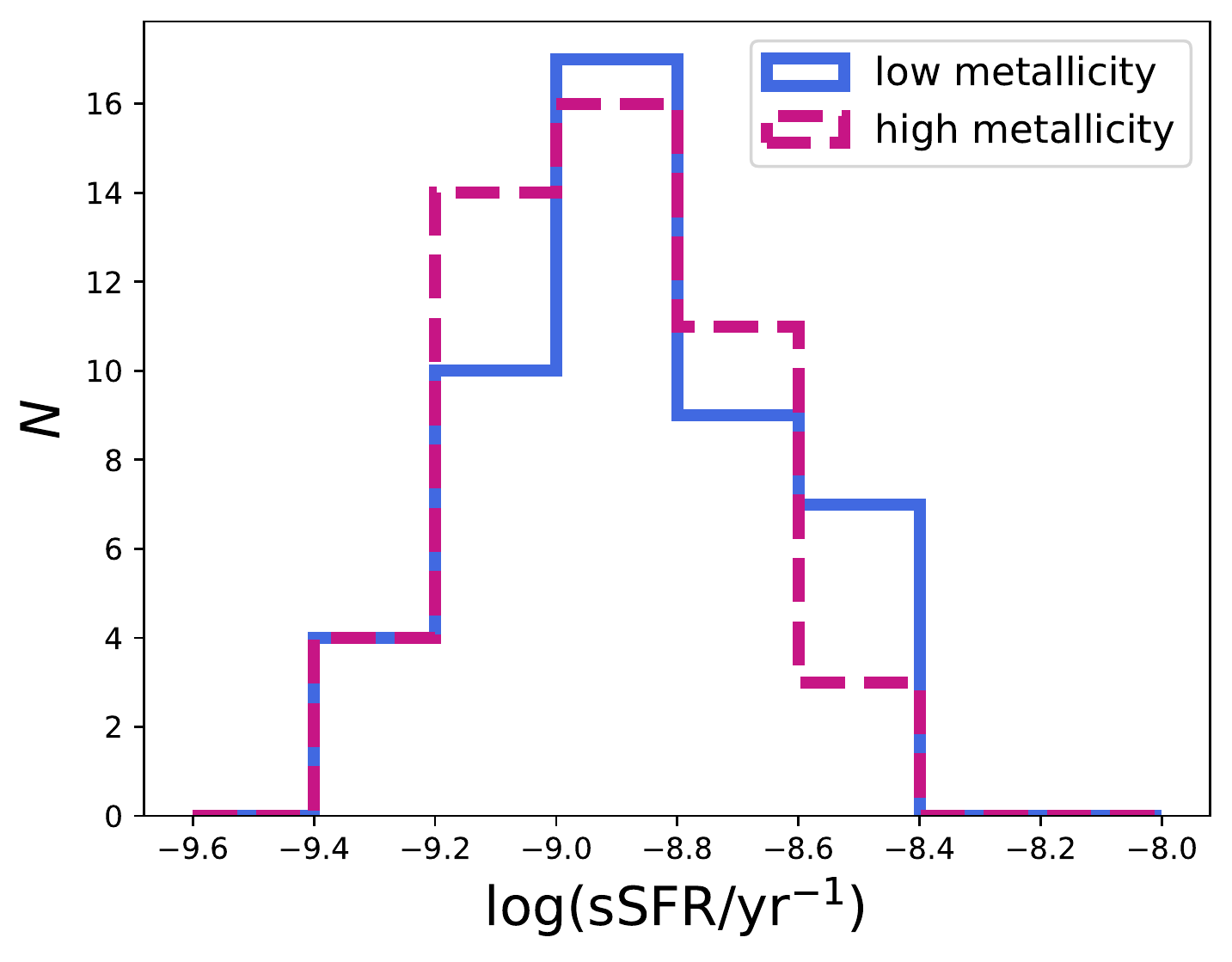} \quad
		\includegraphics[width=.33\textwidth,trim={.2cm 0 0 0},clip]{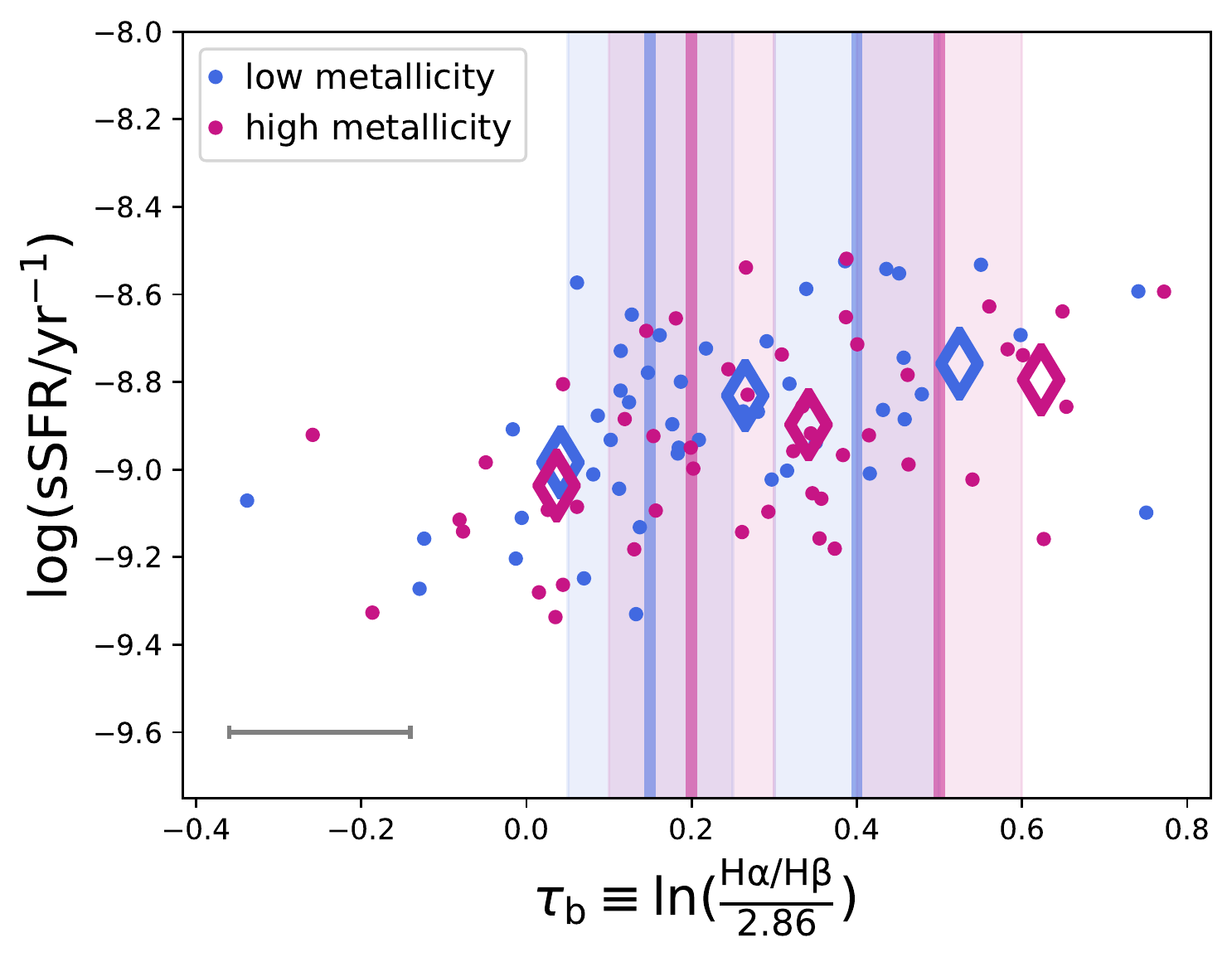} 		
		\caption{{\em Left and Middle:} The Balmer optical depth (left) and specific SFR (middle) distributions of the two high- and low-metallicity subsamples, which are matched to each other within $\Delta(\tau_{\rm b})=0.2$ and $\Delta({\rm \log(sSFR)})=0.2$, respectively.
		{\em Right:} sSFR versus Balmer optical depth in the two metallicity bins. The circles show the individual galaxies, diamonds represent the average values in each {\taub} bin. The vertical lines show the {\taub} boundaries in each metallicity subsample and the shaded regions are the widths at which the {\taub} bin boundaries are perturbed (see Section~\ref{sec:method}). Typical error in {\taub} is shown in the left corner.
		}
		\label{fig:sample-distributions}
\end{figure*}

\subsection{Optical Line Measurements}\label{sec:lines}
Emission-line fluxes are determined by fitting Gaussian functions on top of a linear continuum to MOSFIRE 1D spectra \citep{kriek15,reddy15,freeman19}, and the uncertainties are derived by perturbing the spectrum of each object according to its error spectrum and measuring the standard deviation of the distribution of perturbed line fluxes. A triple Gaussian function was used to fit the {\halpha}$+$[N{\sc ii}]$\lambda6550, 6585$ complex.
Slit-loss corrections are applied by normalizing the spectrum of a ``slit star'' placed on each observing mask to match the 3D-{\em HST} total photometric flux. Additionally, the {\em HST} images of the resolved targets were used to estimate and correct for differential flux loss relative to that of the star described above \citep{reddy15}. Details on line flux measurement and slit loss correction are provided in \citet{kriek15} and \citet{reddy15}.

The {\halpha} and {\hbeta} fluxes are further corrected for underlying stellar Balmer absorption, estimated from the best-fit SED models to the photometry \citep{kriek15,reddy15}. The {\hbeta} absorption fluxes are on average $\sim$15\% of the {\hbeta} observed emission fluxes. The uncertainty in the absorbed fluxes is $\sim$2\% \citep{reddy15}, which is negligible compared to the typical uncertainty of our emission-line fluxes ($\sim$15\%). 

\subsection{SED Fitting and SFR Measurements} \label{sec:sed-sfr}
Throughout this work, the photometry is corrected for contamination from nebular emission lines measured from the MOSDEF rest-frame optical spectra (Sanders et~al., in prep). Corrections are considered for any filter in the 3D-{\em HST} catalog that overlaps with the MOSFIRE $Y$-$J$-$H$-$K$ bands, which includes $z$, $Z$, $Y$, $J$, $H$, $K$, $K_s$, F125W, F140W, F160W, and F850LP filters, whenever available. To determine the emission line correction factor, we create a model spectrum composed of the best-fit Gaussian profile of the emission lines (Section~\ref{sec:lines}) and convolve it with the filter transmission curves. The value of the correction factor depends on the width of the contaminated filter and the emission line equivalent width. In cases that the difference between the original photometry and the emission-line corrected photometry is more than 1$\sigma$ (where $\sigma$ is the photometric error on the observed point), the corrected photometry is adopted. 

To derive stellar masses, and later to normalize the photometry of individual sources to a common base (at rest-frame 5500\,{\AA}, Section~\ref{sec:method}), we use best-fit SED models. The SED fitting is done using the FAST code \citep{kriek09a}, the stellar population model library of \citet{conroy09} for solar stellar metallicity, a \citet{chabrier03} IMF, delayed exponentially declining star-formation history (SFR $\sim t~{\rm exp(}-t/\tau{\rm )}$, where $t$ is age and $\tau$ is the exponential timescale), and the \citet{calzetti00} attenuation curve. The choice of attenuation curve has a statistically insignificant impact on the inferred stellar masses and the normalization at 5500\,{\AA}.

The SFR used to derive sSFR in the following sections is the dust attenuation-corrected {\halpha} SFR ({\sfr}).
We correct {\halpha} luminosity for attenuation using the Balmer decrement and the \citet{cardelli89} MW extinction curve \citep[see][]{reddy15,shivaei15b,shivaei16}. Assuming that the H{\sc ii} regions follow a foreground dust screen model, the MW curve is preferable for the extinction correction of nebular lines, as it is derived based on the line-of-sight measurements of H{\sc ii} regions \citep{calzetti94,wild11b,liu13,salim20} and is shown to be an accurate representation of nebular attenuation curve at $z\sim 2$ (Reddy et~al., in prep). Multiple studies have also shown a good agreement between dust-corrected SFR({\halpha}) using the MW curve and IR-inferred SFRs \citep[e.g.,][]{flores04,shivaei16}. The dust-corrected {\halpha} luminosities are converted to SFR based on the \citet{hao11} conversion, modified for a Chabrier IMF.

\subsection{Sample}
\label{sec:sample}
We start with a sample of 271 galaxies from the MOSDEF survey with {\halpha}, {\hbeta}, and [N{\sc ii}] detections. To determine the attenuation curve, we must ensure that the intrinsic (i.e., dust free) SEDs of the galaxies have similar shapes. Therefore, avoiding galaxies that are optically thick in the UV or in optical nebular lines, and those with very young ages whose intrinsic UV slope is a strong function of stellar population age, is necessary. 
We impose the following criteria on the parent MOSDEF sample to obtain a clean sample for our analysis:
\begin{itemize}
\item Robust ($>3\sigma$) detection of {\halpha} and {\hbeta} emission lines;
\item Removal of targets with AGN contamination based on X-ray emission, IRAC colors, and/or [N{\sc ii}]/{\halpha} ratios \citep[for details on MOSDEF AGN identification see][]{coil15,azadi17,azadi18,leung19};
\item Removal of ULIRGs (L(IR) $>10^{12}\,${\lsun}) to ensure the sample is clean of heavily dust-obscured objects. We use {\em Spitzer}/MIPS 24{\um} flux densities and the redshift-dependent conversion of \citet{rujopakarn13}, to estimate total IR luminosities of individual objects;
\item Removal of targets with [O{\sc iii}] and {\halpha} line equivalent widths greater than 630\,\AA~and 1600\,\AA, respectively, to avoid galaxies with ages younger than 100\,Myr \citep[the limits are determined based on the results of][]{reddy18b};
\item Removal of quiescent targets using UVJ criteria \citep{zick18} and those with low sSFRs ($\log({\rm sSFR/yr^{-1}})<-9.6$);
\item Removal of targets with Balmer optical depth of \taub $>1$ (Equation~\ref{eq:taub} below) to avoid optically-thick objects. \citet{shivaei16} have shown that {\sfr} agrees well with the bolometric UV-to-IR SFR up to $\sim 100$\msun/yr in IR-detected galaxies. This SFR limit corresponds to {\taub} $\sim 1$, based on a linear regression fit to {\sfr} versus {\taub} in our current sample;
\item Robust ($>3\sigma$) detection of [N{\sc ii}]$\lambda6585$ for metallicity measurements, see below, as the main goal of this work is to evaluate the change of the attenuation curve as a function of metallicity. 
\end{itemize}
These criteria result in a sample of 218 galaxies at $z= 1.4-2.6$. 

\subsubsection{Sample with Metallicity Measurements} \label{sec:metal-sample}

The [N{\sc ii}] and {\halpha} lines are used to estimate the gas-phase metallicity (oxygen abundances, $12+\log{(\rm O/H)}$), assuming the \citet{pp04} empirical calibration \citep{sanders15,shapley15}. The [N{\sc ii}]/{\halpha} ratio is shown to be a good tracer of metallicity in the local universe for the range of metallicities that we trace \citep[$12+\log{(\rm O/H)}\sim 8.0-8.8$;][]{pp04,marino13}, but at high redshifts, the ratio has uncertainties as a diagnostic of oxygen abundance due to N/O variations and other uncertainties regarding the physical conditions in H{\sc ii} regions \citep[e.g.,][]{shapley15,strom17,bian18}. Therefore, the absolute metallicity values should be taken with caution. However, using [N{\sc ii}]/{\halpha} to sort galaxies by metallicity, as in this study, is robust, as the ratio correlates well with other metallicity indicators that use oxygen lines \citep[e.g.,][]{steidel14,sanders15}.

We divide the sample into two metallicity bins at $12+\log{(\rm O/H)}=8.5$, hereafter low- and high-metallicity bins. 
The oxygen abundance of $12+\log{(\rm O/H)}=8.5$ is the [N{\sc ii}]/{\halpha} metallicity limit below which the mid-IR aromatic band (PAH) intensity at rest-frame 7.7\,{\um} (defined as the ratio of 7.7\,{\um} luminosity to total SFR) decreases significantly in the MOSDEF sample, as shown in \citet{shivaei17}.
As PAHs are suggested as one of the main carriers of the UV extinction bump \citep{stecher65,draine03,weingartner01a,li01}, we divide the sample at this metallicity limit to look for variations in the bump strength.
There are 83 and 135 galaxies in the low- and high-metallicity bins.

We further impose three additional criteria to avoid biases due to different optical depths and intrinsic SEDs in the two metallicity subsamples and within each metallicity subsample when divided into bins of Balmer optical depth ($\tau_{\rm b}\equiv \tau_{\rm H\alpha}-\tau_{\rm H\beta}$, see Section~\ref{sec:method}):

\begin{itemize}

\item In the absence of short-term star formation variations, large differences in the sSFR may indicate differences in the shape of the SED. Therefore, we limit the sSFRs to $-8.5<\log({\rm sSFR/yr^{-1}})<-9.4$ to avoid large variations in the sSFR of galaxies as a function of {\taub} in each metallicity subsample (see the discussion in the Appendix).
This criterion helps with having similar intrinsic stellar populations within the {\taub} bins and mitigating the uncertainties associated with the change of intrinsic SED as a function of {\taub}  \citep{chevallard13}. We estimate the systematic uncertainty within our sample associated with the SFR stochasticity and variations in the average sSFRs in Section~\ref{sec:slope-unc} and the Appendix, showing they do not affect our results. Figure~\ref{fig:sample-distributions}-right shows sSFR versus {\taub} in the two metallicity subsamples;

\item To control for the variations of the intrinsic stellar populations between the two metallicity subsamples, we match the two subsamples in their sSFR distribution, as follows. For each high-metallicity galaxy with a given sSFR, a galaxy from the low-metallicity sample with a sSFR within 0.2\,dex of that of the high-metallicity galaxy is selected (see Section~\ref{sec:method}). The K-S statistic of the two sSFR distributions is 0.14 with a p-value of 0.72, consistent with the hypothesis that the sSFR distributions of the two samples are the same. The distributions are shown in Figure~\ref{fig:sample-distributions}-middle;

\item In the same way as matching the sSFR distributions, the {\taub} distributions of the two metallicity subsamples are also matched within $\Delta{\rm \tau_b}=0.2$ (K-S statistic of 0.14 with a high p-value of 0.71, consistent with the hypothesis that the {\taub} distributions of the two samples are the same), to control for the effect of different optical depth distributions in the low- and high-metallicity attenuation curves, as it is known that the shape of the attenuation curve varies with the optical depth \citep{chevallard13,salmon16,salim18}. The distributions are shown in Figure~\ref{fig:sample-distributions}-left. 

\end{itemize}
The final sample has 48 objects in each metallicity bin. 

\begin{figure}[t]
		\includegraphics[width=.48\textwidth,trim={.2cm 0 0 0},clip]{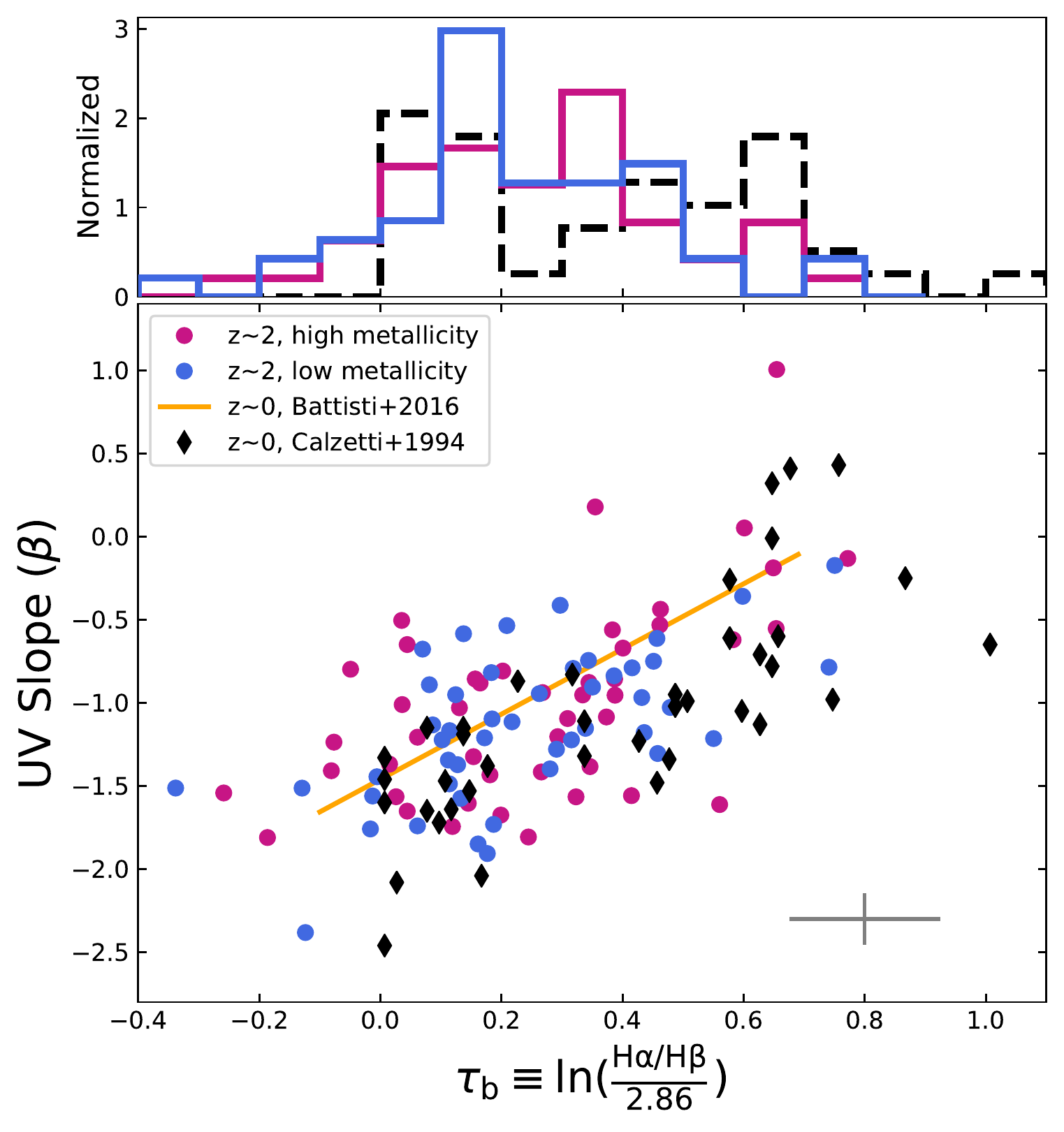}
		\caption{UV continuum slope versus Balmer optical depth for targets in our sample. Galaxies with metallicity measurements from [N{\sc ii}]/{\halpha} line ratios \citep{pp04,sanders18} are shown with blue and magenta colors, separated at $12+\log({\rm O/H})=8.5$ into low- and high-metallicity bins, respectively.  The orange solid line is the $\beta-${\taub} fit of the $z\sim 0$ sample of \citet{battisti16}. The local starburst sample of \citet{calzetti94} is shown with black diamonds. The normalized {\taub} distributions of the low-metallicity sample (blue), high-metallicity sample (magenta), and the Calzetti sample (dashed black) are shown in the upper panel. 
		}
		\label{fig:taub-beta}
\end{figure}

The UV continuum slope ($\beta$ in $f_{\lambda}\propto \lambda^{\beta}$ at $\lambda=1260–2600$\,\AA) is an indication of the stellar reddening for typical star-forming galaxies \citep{meurer99}. The relation between {\taub} and $\beta$ in the two metallicity samples is shown in Figure~\ref{fig:taub-beta}, where $\beta$ is calculated by fitting a power-law function to the best-fit SED models (Section~\ref{sec:sed-sfr}), $f_{\lambda}\propto \lambda^{\beta}$, at $\lambda \sim 1260-2600$\,{\AA}. 
The good correlation between $\beta$ and {\taub} (Pearson correlation coefficient of $\sim 0.6$ with p-value of $\lesssim 10^{-5}$) and similar trends with the original \citet{calzetti00} and \citet{battisti16} local samples show the validity of the adopted methodology to derive the stellar attenuation curve, as discussed in Section~\ref{sec:method}. 

\begin{figure*}[th]
	\centering
		\includegraphics[width=.325\textwidth,trim={0 0 0 0},clip]{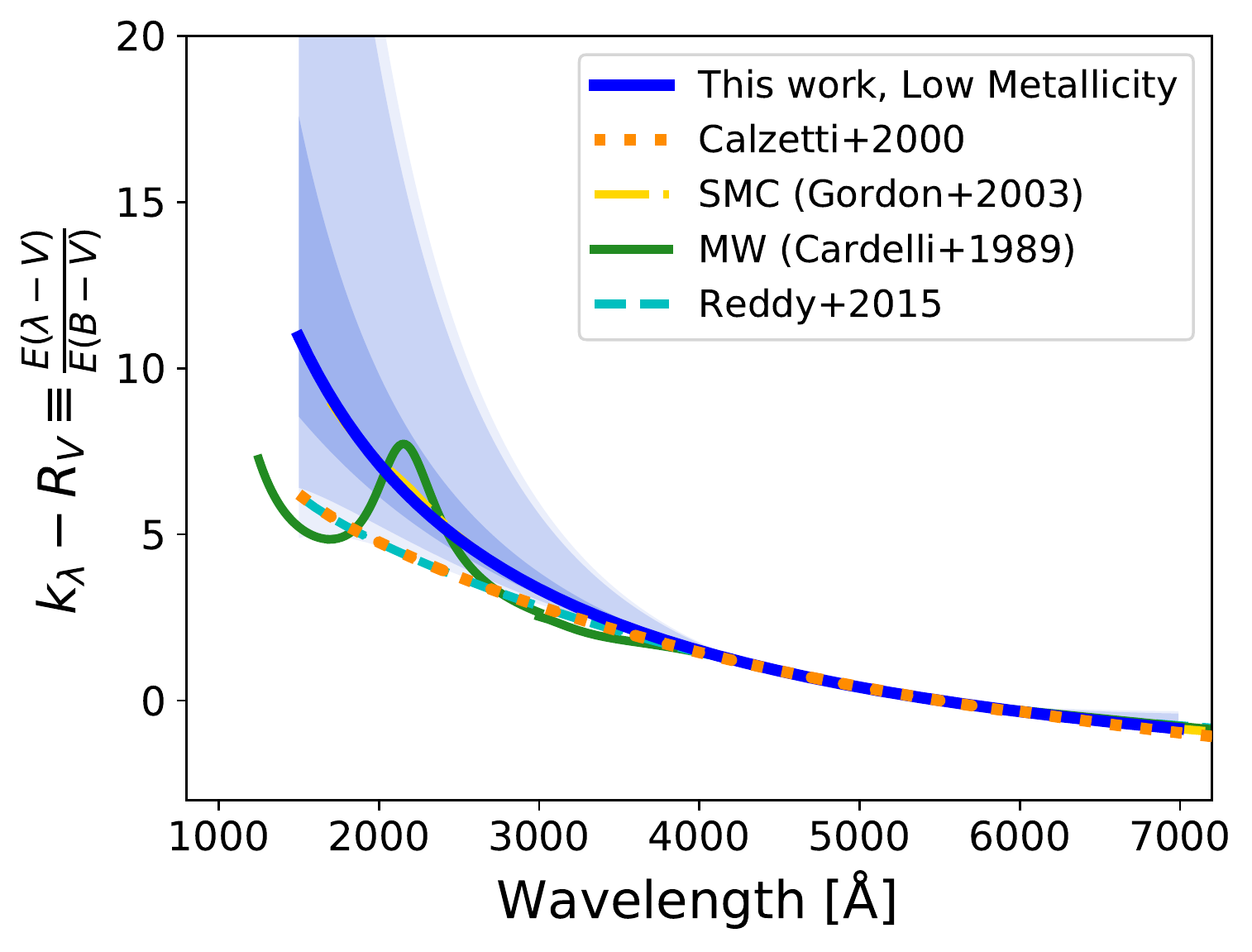}\quad
		\includegraphics[width=.325\textwidth,trim={0 0 0 0},clip]{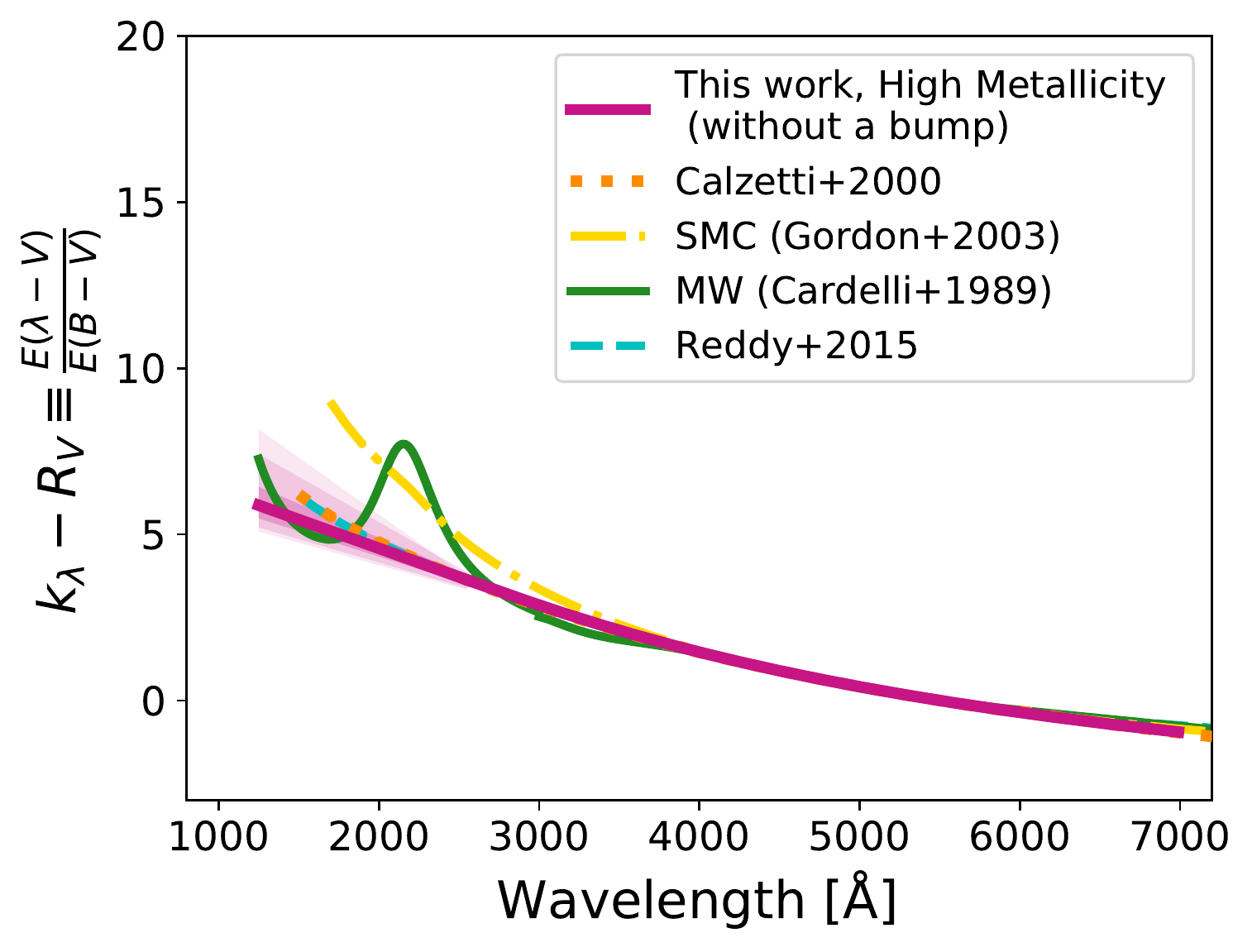}
		\includegraphics[width=.325\textwidth,trim={0 0 0 0},clip]{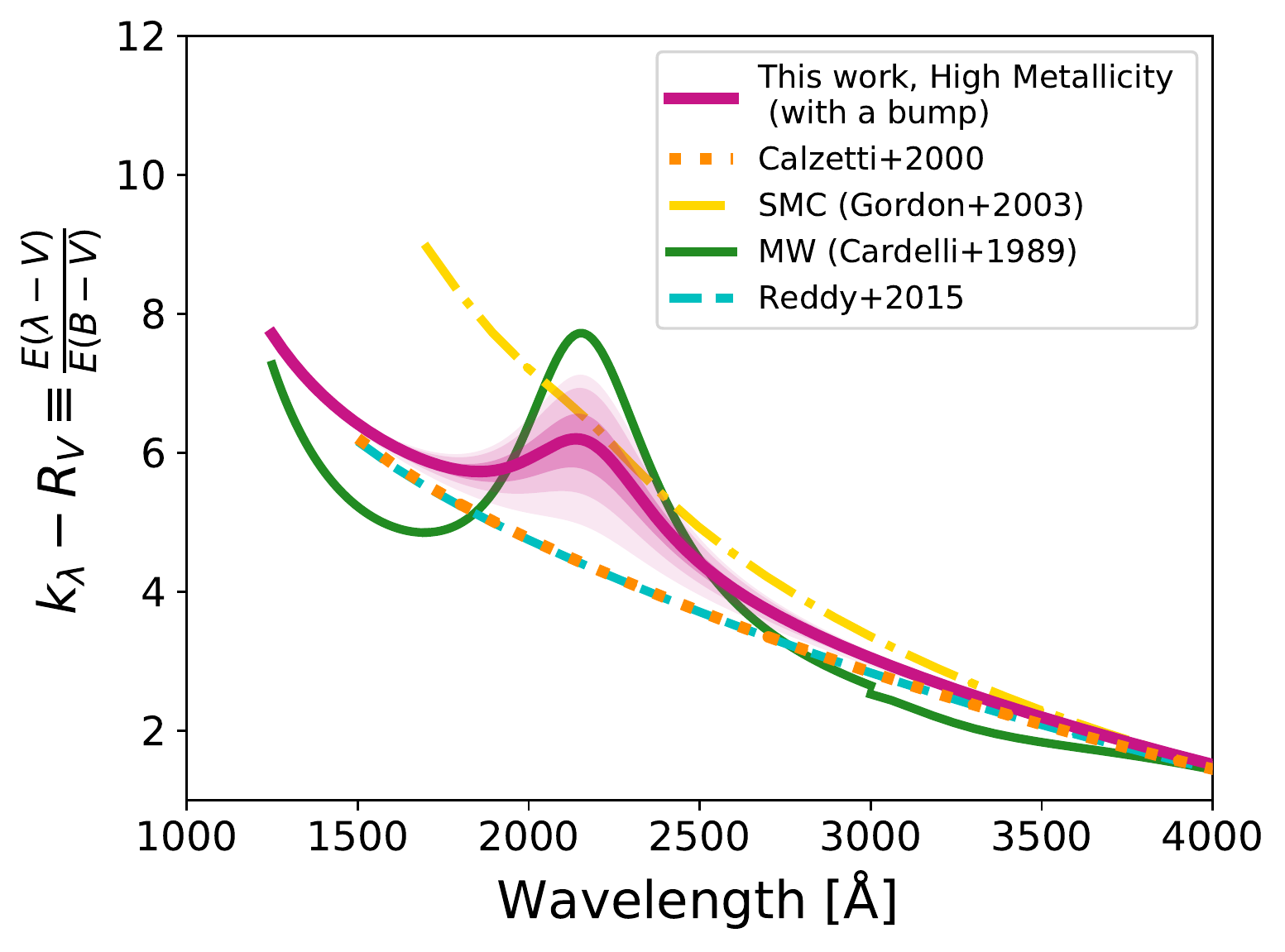}\quad
		\caption{Stellar attenuation curves at $z=1.4-2.6$  for the two samples with gas-phase metallicities below and above $12+\log({\rm O/H})=8.5$. The shaded regions show 1, 2, and 3 standard deviations from the mean curves. {\em Left:} The low-metallicity curve is similar to the SMC curve with a steep UV rise. {\em Middle:} the UV slope of the high-metallicity curve is identical to that of the R15 and C00 starburst curves. This fit does not include a Drude profile for the UV extinction bump. 
		{\em Right:} The bump in the high-metallicity curve assuming the underlying shape of the C00 starburst curve. The bump strength is about 0.5 of the MW bump strength.
		}
		\label{fig:att-metal}
\end{figure*} 

\section{Results}
\label{sec:curves}
It is expected that the physical properties of dust grains (e.g., dust grain composition), and as a result the shape of the attenuation curve, vary with ISM gas abundances. \citet{shivaei17} have shown that the emission intensity of the PAH dust grains in the mid-IR, suggested as the origin of the UV extinction bump \citep{stecher65,draine03,weingartner01a,li01}, strongly depends on the gas metallicity at $z\sim 2$ \citep[consistent with similar studies at lower redshifts, e.g.,][]{engelbracht05,draine07b,smith07,marble10}. 
Motivated by such studies, we explore the variations of the shape of the curve (UV rise and the UV extinction bump) with metallicity in this section.
We use the methodology described in Section~\ref{sec:method} to derive the attenuation curves ($fQ_{\rm eff}\equiv k_{\lambda}-R_V$) in two metallicity bins, separated at $12+\log{(\rm O/H)}=8.5$, as defined in Section~\ref{sec:sample}.

\subsection{Variations in the curve slope}
\label{sec:slope}

The left and middle panels in Figure~\ref{fig:att-metal} show the attenuation curves in the two metallicity bins, fit with a second-order polynomial without a Drude profile (Equation~\ref{eq:qeff}; the UV extinction bump will be investigated in the next section). The shaded regions show the 68, 95, and 99.7 percentile bounds from our Monto Carlo analysis (Section~\ref{sec:method}). 

The low-metallicity curve shows a steep UV rise, similar to that of the SMC curve from \citet{gordon03}. Nonetheless, the curve is consistent with the C00 and the \citet{reddy15} (hereafter, R15) curves at 3$\sigma$. On the other hand, the high-metallicity curve is identical to the C00 and R15 curves.\footnote{As we use a quadratic function to fit the curves, the high-metallicity fit had a turnover to low attenuation curve values at $\lambda \lesssim 2000$\,{\AA}, which is not physically motivated, and hence corrected for by linearly extrapolating the quadratic curve to short wavelengths (Figure~\ref{fig:att-metal}-middle).}
Our result of a steeper attenuation curve at low metallicities is also applicable to galaxies with low masses, given the correlation between stellar mass and gas-phase metallicity. This result is in agreement with previous studies showing that steeper attenuation curves are favored in low-mass galaxies at high redshifts \citep{reddy18a,alvarez19}.

We note that the R15 curve was derived using a sample that partially overlaps with ours, separated into two sSFR bins. Using the same sample as in R15, we find the exact same results, except for a stronger indication of the UV bump as discussed below. It is when the samples are divided into metallicity bins that the low-metallicity curve starts to deviate in the UV (bluewards of 2500\,{\AA}) from the shallower curve of the high-metallicity bin and the R15 curve. 

\subsubsection{Slope Uncertainties} \label{sec:slope-unc}

The uncertainty intervals shown in Figure~\ref{fig:att-metal} encompass measurement uncertainties, and uncertainties associated with sample selection and {\taub} bin variations (see Section~\ref{sec:method}). The $1\sigma$ relative uncertainties, $\Delta Q_{\rm eff}/Q_{\rm eff}$, at $\lambda=1500$\,{\AA} of the low- and high-metallicity curves are ($+48\%,-23\%$) and ($+9\%,-7\%$), respectively. The measurement uncertainties associated with the flux measurement errors are ($+20\%,-15\%$) and ($+7\%,-5\%$) for the low- and high-metallicity curves, respectively. As is evident, the uncertainty in the high-metallicity curve is dominated by the measurement uncertainties. On the other hand, the uncertainty of the low-metallicity curve is much larger than the measurement uncertainties, which suggests a larger intrinsic scatter in the attenuation curves of low-metallicity galaxies compared to those of the high-metallicity galaxies.

The other source of uncertainty in the Balmer optical depth method used here is associated with the variations in the {\em intrinsic} SEDs of galaxies used to derive the attenuation curve.
\citet{calzetti94} pioneered determination of the attenuation curve using Balmer optical depths by appealing to the similarity of galaxy SEDs in the $1200-2600$\,{\AA} range, assuming a constant rate of star formation. They pointed out that even a young, instantaneous star formation burst has a spectral slope in this range deviating by only 13\%, within the errors of their study. Our study depends on a similar assumption about the UV SEDs; in the appendix, we test its validity given current understanding of the MOSDEF galaxies. 
There we demonstrate that the average slopes used in our analysis are not influenced by SED variations significantly compared with the other errors in the results.
Furthermore, limits we place on the sSFR in Section~\ref{sec:metal-sample} make the sSFR distribution between the two metallicity samples similar. Therefore, any potential uncertainty would affect the two metallicity curves similarly, and would not affect the results in terms of comparing the attenuation curve of the low- and high- metallicity samples.

In summary, the average attenuation curve for our low metallicity sample agrees well with the prototypical curve for low metallicity galaxies from the SMC. Based on the errors induced by measurement uncertainties alone, it is unlikely that the Calzetti/high-metallicity curve applies to the low metallicity sample on average. The formal significance of this result of $\sim 3\sigma$ is probably understated, because it is likely that the derived uncertainty is amplified by intrinsic scatter in the curve among the low-metallicity galaxies. 

\subsection{Variations in the UV extinction bump}
\label{sec:bump}

In this section, we fix the shape of the attenuation curve to those derived in the previous section (Section~\ref{sec:slope}) and investigate the presence of a UV extinction bump in the data. We re-fit the effective attenuation curves, this time by adding a Drude function to the best-fit polynomials from the previous section (Equation~\ref{eq:drude}). 

\begin{figure*}[th]
	\centering
		\includegraphics[width=	.33\textwidth,trim={0 0 0 0},clip]{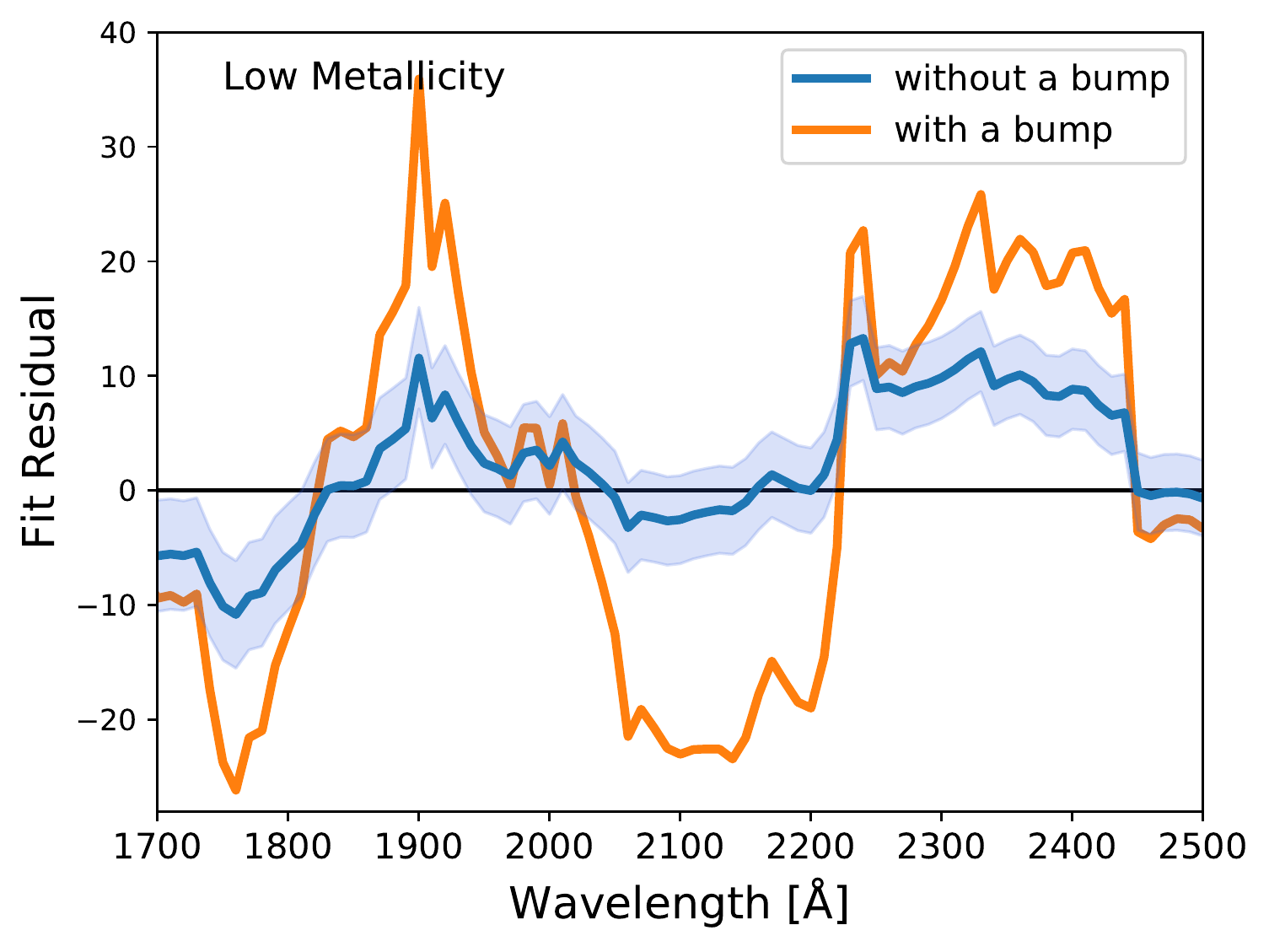}\quad
		\includegraphics[width=.33\textwidth,trim={0 0 0 0},clip]{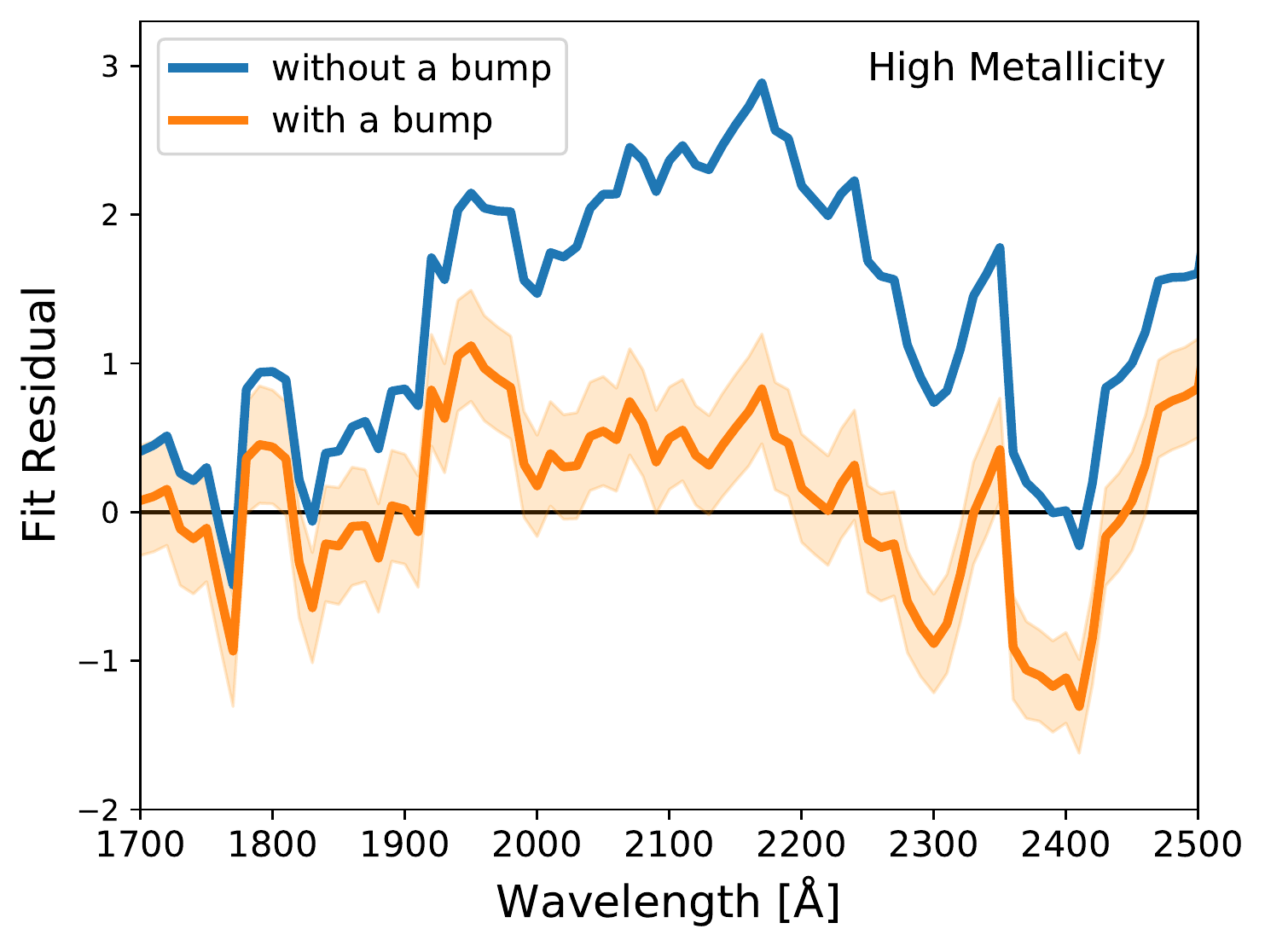}\quad
		\caption{The residual of the best-fit effective attenuation curve ($fQ_{\rm eff}-$model) with (orange) and without (blue) a Drude profile UV bump for the low- and high-metallicity samples in the left and right panels, respectively. The shaded regions reflect the error in the fit given the measurement uncertainties of the input flux densities. The black horizontal line shows zero residual. The high-metallicity sample is best fit with a UV bump, while no bump is indicated for the low-metallicity sample by $>3\sigma$.
		}
		\label{fig:bump-metal}
\end{figure*}

In Figure~\ref{fig:bump-metal}, we show the fit residuals (data$-$best-fit model) with and without a UV bump for the two metallicity bins. The high-metallicity curve is significantly improved when a UV bump is added, given that the residuals of the fit including a bump are consistent with zero within 2$\sigma$. On the other hand, a bump with similar strength as the high-metallicity bump is ruled out by $>3\sigma$ in the low-metallicity curve. The low-metallicity fit without a bump shows residuals consistent with zero within 1$\sigma$.

To constrain the strength of the UV bump in the high-metallicity bin, we fit the effective attenuation curve of all galaxies with $12+\log{(\rm O/H)}>8.5$ in our sample with the C00 effective curve ($k_{\lambda}-R_V$) and a Drude profile, leaving the amplitude of the bump as a free parameter. For this part, we remove the previous restrictions on the sSFR and {\taub} to be matched to the low-metallicity distributions. The total sample consists of 83 galaxies. We find that the bump amplitude for the high-metallicity sample is $E_b=1.53^{+0.33}_{-0.37}$, shown in Figure~\ref{fig:att-metal}-right. For comparison, the MW bump amplitude is $E_b=3.3$ \citep{fitzpatrick99}. The relative strength found here agrees with that reported by \citet{buat11} and is not significantly different from the reports by \citet{noll09}, \citet{kriek13}, \citet{salim18}, and \citet{battisti20}.

\begin{figure*}[t]
	\centering
		\includegraphics[width=.55\textwidth,trim={0 0 0 0},clip]{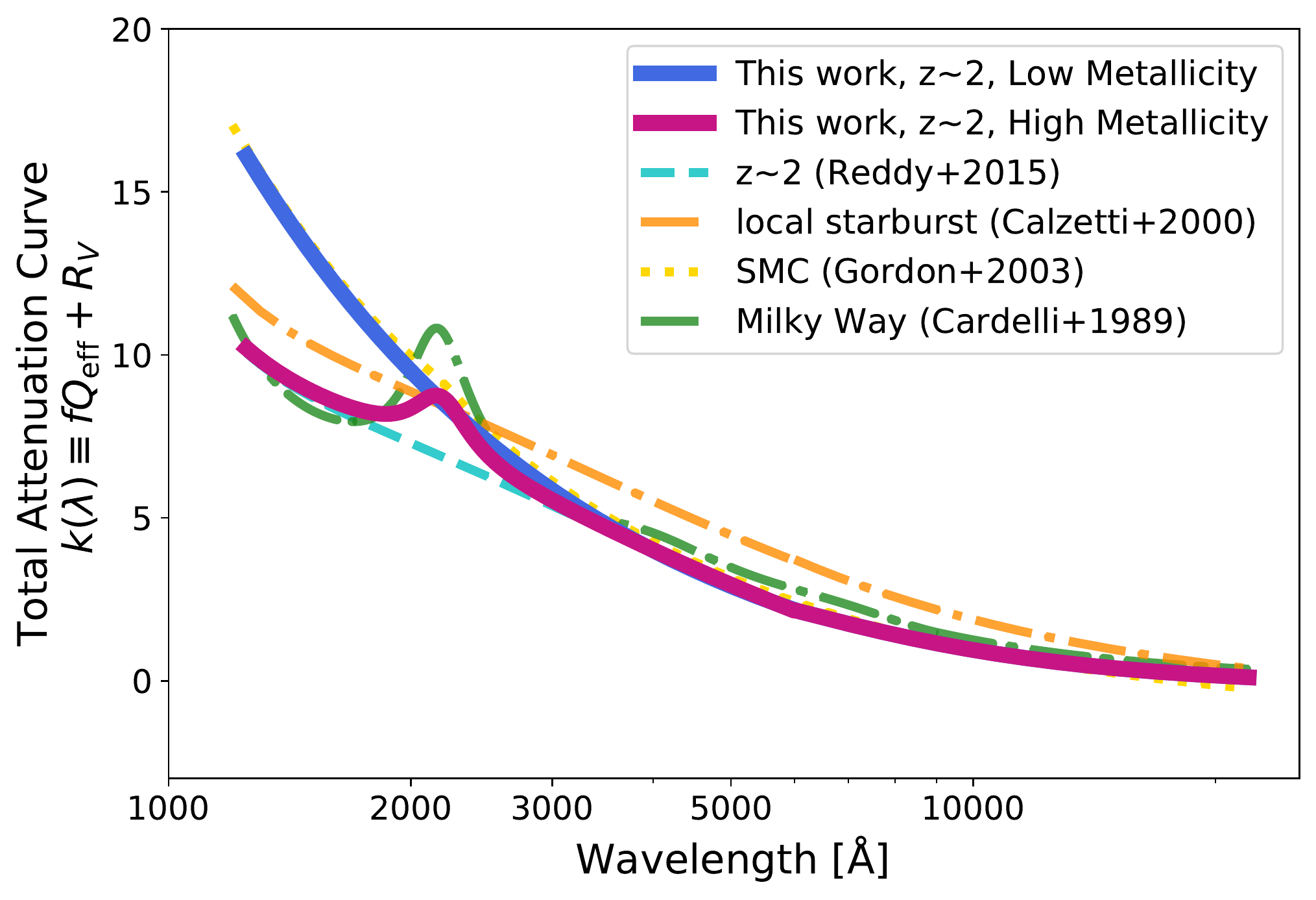}
		\caption{The total attenuation curves for the two metallicity subsamples, separated at $12+\log{\rm (O/H)}=8.5$, corresponding to $M_*\sim 10^{10.4}$\,{\msun}. The high-metallicity curve (violet) is estimated as the \citet{calzetti00} effective curve with a UV bump with $E_b=1.53$ (Equation~\ref{eq:drude}) at $\lambda<6000$\,{\AA} and the \citet{reddy15} curve at $\lambda>6000$\,{\AA}. The low-metallicity curve is estimated as the \citet{gordon03} SMC curve without a UV bump at $\lambda<6000$\,{\AA} and the \citet{reddy15} curve at $\lambda>6000$\,{\AA}. The \citet{gordon03} SMC curve lies largely underneath our low-metallicity curve.
		In both the high- and low-metallicity curves, $R_V=2.505$.
				}
		\label{fig:totalcurves}
\end{figure*}

\subsection{Total attenuation curves}
\label{sec:total}
The effective attenuation curves ($k_{\lambda}-R_V$) derived in this work can be approximated by the SMC curve for the low-metallicity sample and the C00 starburst curve with a UV extinction bump with the strength of $E_b=1.53$ (Equation~\ref{eq:drude}) for the high-metallicity sample. 

To compare our curves to the total attenuation curves ($k_{\lambda}$) from other studies, we need to assume an $R_V$ value and a functional form for the long-wavelength ($\lambda > 0.60$\,{\um}) part of the curve. The long-wavelength shape of the curve is dominated by diffuse dust in the ISM and less affected by the dust grain properties and dust-star geometry effects around massive stars. Similarly, in the absence of IR data, the best way of determining $R_V$ is by extrapolating the curve to long wavelengths \footnote{ The most direct way of determining $R_V$ is by using IR data and implementing energy balance between the absorbed and  re-emitted light \citep{calzetti00}. However, given the large uncertainties in current IR measurements of high-redshift galaxies (using {\em Spitzer} and {\em Herschel}), inferring $R_V$ is still a challenging task and beyond the scope of this paper.}.
As the sample used in this study and that of R15 are from the same MOSDEF parent sample and overlap with each other to a large extent, we take the shape of the long-wavelength ($\lambda > 0.60$\,{\um}) curve and $R_V=2.505$ of the R15 curve to calculate the total attenuation curve. The $R_V$ is determined by setting a limit condition of $k_{\lambda}=0$ at $\lambda=2.85$\,{\um}. Extrapolating the curve to $k_{\lambda \rightarrow \infty}$ results in $R_V=2.672$, and extrapolating it to $\lambda= \infty$ with a functional form of $1/\lambda$ results in $R_V=2.942$ \citep{reddy15}. The inconsistency among different values reflects the uncertainty in determining $R_V$ by extrapolating the attenuation curve to near-IR wavelengths, as the curve is not constrained at $\lambda>2$\,{\um}. For the rest of this work, we adopt $R_V=2.505$, because it requires the least amount of extrapolation. The total attenuation curves, shown in Figure~\ref{fig:totalcurves}, are:

\begin{equation} \label{eq:klam}
\begin{split}
k({\lambda})  = -2.659&\,(-2.156+1.509\,x-0.198\,x^2+0.011\,x^3) \\
 & + 1.53\, \frac{x^2}{(x^2-x_0^2)^2+x^2 \gamma^2}  + 2.505,\\
 & 0.12\leq \lambda< 0.60\,{\rm \mu m},~{\rm high~metallicity/mass;}\\
  = -4.116 &+ 2.264\,x  + 2.505,\\
 & 0.12\leq \lambda<0.60\,{\rm \mu m},~{\rm low~metallicity/mass;}\\
  =  -2.672 &- 0.010\,x + 1.532\,x^2 - 0.412\,x^3 + 2.505,\\
 & 0.60\leq \lambda<2.85\,{\rm \mu m},
\end{split}
\end{equation}
\noindent where $x=1/\lambda$ in {\um}$^{-1}$, and the central wavelength ($x_0$) and width ($\gamma$) of the UV extinction bump are the MW values of 4.59{\um}$^{-1}$ and 0.922{\um}$^{-1}$, adopted from \citet{fm07}.

Owing to the mass-metallicity relation \citep[e.g.,][]{tremonti04}, we may attribute our low- and high-metallicity curves to low- and high-mass galaxies, respectively. The variation of the attenuation curve with mass has been a controversial topic in the literature. While \citet{salmon16} did not find any correlation between stellar mass and the slope of the curve or the strength of the UV bump, \citet{zeimann15} concluded that the UV slope of the attenuation curve increases with increasing mass (but did not find a correlation with the UV bump strength), and attributed the trend to dust-star geometry. The slopes found in \citet{zeimann15}, over the mass range of $\log(M_*/M_{\odot})=7.2-10.2$, were consistently lower than the C00 slope (and the slopes found in this study).  On the other hand, \citet{reddy18a} found that the slope of the curve increases with decreasing mass. In this work, extrapolating from the mass-metallicity relation, we see a similar trend to that of \citet{reddy18a}, such that the high-mass galaxies have shallower attenuation curves (and stronger UV bumps) compared to low-mass galaxies. We mainly attribute this trend to the differences in dust grain properties of low- and high-metallicity galaxies (discussion in Section~\ref{sec:imp-dust}). 

The metallicity division of our two samples, $12+\log(\rm{O/H})=8.5$, corresponds to a stellar mass of $M_*\sim 10^{10.4}$\,{\msun}, according to the MOSDEF mass-metallicity relation for the same [N{\sc ii}]/{\halpha} metallicity indicator from Table 1 of \citet{sanders18}. However, we note that the cut in metallicity does not correspond to a clean cut in stellar mass, owing to the large scatter in the mass-metallicity relation \citep[see Figure 2 in][]{sanders18}. 

\section{Discussion and Implications}
\label{sec:implications}

In this section, we review some of the implications of the attenuation curves derived in this study. First, in Section~\ref{sec:imp-dust}, we discuss how different dust grain properties and dust-star geometry may explain the observed variations in the UV attenuation curve as a function of metallicity. We compare the SFRs calculated using our metallicity-dependent curves with {\sfr} in Section~\ref{sec:sfr}. Section~\ref{sec:beta} shows how a UV extinction bump may affect the photometry of high-redshift galaxies, and hence the measurements of the UV continuum slope. Lastly, in Section~\ref{sec:ebmv}, we present the comparison of stellar reddening ($E(B-V)$), derived from the UV slope and assuming our new attenuation curves, with gas reddening, derived from Balmer nebular lines.

\subsection{Dust grains and geometry} \label{sec:imp-dust}
\paragraph{The UV slope of the curve}
A steep rise in the UV extinction can be produced by varying the dust grain size distribution and/or grain composition. Small grains are shown to produce steep extinction curves \citep{hou17}, hence an environment with shocks or intense radiation that shatters dust grains and produces small grains can result in a steep UV rise \citep{gordon97}.
The other possibility is a lower fraction of small carbonaceous to small silicate grains \citep{weingartner01a}, which may result from delayed carbon enrichment from AGB stars \citep{zhukovska13} or more efficient destruction of small carbonaceous grains relative to small silicate grains by supernovae \citep{hou16}. 
Such environments may be more common in the turbulent ISM of actively star-forming galaxies at $z\sim 2$ and in lower metallicity galaxies that have preferentially younger ages and also more intense radiation \citep{sanders16a,strom18}. Also, the molecular fraction (ratio of molecular to total gas mass) is a function of metallicity, such that the low-metallicity galaxies have fewer molecular clouds \citep[e.g.,][]{krumholz09,fumagalli10}. Therefore, the decreased gas and dust shielding from UV radiation in low-metallicity galaxies can provide a more favorable environment for shattering large grains and hence, a steeper UV attenuation curve.

Radiative transfer effects, such as a higher dust column density, turbulent or clumpy ISM (instead of homogeneous), and a geometry with a mixture of stars and dust (as opposed to a screen/shell of dust), generally result in flatter attenuation curves and weaker bump strengths \citep{wittgordon00,seon16,popping17}. 
However, an age-dependent attenuation, in which young stars are embedded in their more-than-average dusty birth clouds and are more attenuated than older stars, can explain a steep UV attenuation curve \citep{inoue05,panuzzo07,tress18,tress19}. 
An age-dependent attenuation model suggests that older stars, dominating the emission at long wavelengths, are subject to less attenuation compared to younger stars that dominate the emission at short wavelengths, resulting in a steep attenuation curve. Similar trends have also been seen at $z\sim 2$, where young galaxies show a lower IR-to-UV luminosity at a given UV continuum slope, suggesting a steeper attenuation curve \citep{reddy10,reddy12a,reddy18a} or the stacks of young galaxies indicate a better agreement with IR+UV SFRs if a steeper than C00 curve is used to correct the observed UV SFRs \citep{shivaei15a}.
\citet{inoue05} also attributed a steep attenuation curve slope to a decreased optical thickness in galaxies with lower dust-to-gas mass ratios. Therefore, we expect that given our low metallicity galaxies are younger\footnote{Average ages of our low- and high-metallicity samples, derived from SED fitting (Section~\ref{sec:sed-sfr}), are $\log({\rm age/yr})= 8.67\pm 0.02$ and $8.87\pm 0.03$, respectively, with a large scatter of $\sim 0.3$\,dex. We note that the ages inferred from SED fitting assuming rising star-formation histories are ambiguous \citep[e.g.,][]{reddy12b}.} \citep{gallazzi05} and expected to have lower dust-to-gas ratios compared to higher metallicity galaxies \citep{remyruyer14}, they would show a steeper UV attenuation curve. In this study, we cannot robustly distinguish the effect of different grain properties from that of age-dependent geometries on the slope of the curve.

\paragraph{The UV extinction bump}

Our data show an unambiguous signature of a UV bump in the attenuation curve of the high-metallicity sample and rule out a bump of similar strength in the low-metallicity curve by $>3\sigma$. 
Theoretically, the strength of the bump in an attenuation curve (assuming the underlying extinction curve has a bump) correlates with the UV slope of the curve, such that the radiative transfer effects that suppress the bump (e.g., high optical depth or clumpy ISM) also make the curve flatter \citep{wittgordon00,seon16}. This effect has also been indicated in observations \citep{kriek13,salim18}. 
However, we see the opposite of this trend: the high-metallicity curve is shallower but has a stronger bump compared to the low-metallicity curve. This result suggests that the absence of a significant bump in the steep low-metallicity curve is a reflection of dust grain properties, rather than radiative transfer effects (which includes the dust-star geometry effects).
Moreover, the bump strength observed in the high-metallicity curve is likely a lower limit on the intrinsic bump strength in the extinction curve of high-metallicity galaxies, owing to the same argument that the observed bump is likely suppressed due to geometrical effects.

A significant difference between the two metallicity samples is the change of the intensity of the mid-IR aromatic band at 7.7\,{\um}, attributed to emission from PAH molecules. The 7.7\,{\um} luminosity to total SFR ratio changes from $\sim 35\pm 9$ (in units of $10^8\,L_{\odot}/M_{\odot}{\rm yr^{-1}}$) in the high-metallicity sample to $\sim 6\pm 1$ in the low metallicity sample (see Table 1 and Figure 3 of \citealt{shivaei17}). The decrease in the PAH intensity can be caused by either preferential destruction of PAHs by the harder ionizing radiation of low metallicity environments or a delayed enrichment of PAHs in young (and low metallicity) galaxies \citep[e.g.,][]{dwek05,draine07b,smith07,galliano08}.
The trend between the UV bump strength and metallicity in this work combined with the trend of the PAH intensity with metallicity from \citet{shivaei17} may suggest a correlation between the strength of the UV extinction bump and the PAH emission intensity. PAHs have been suggested as the origin of the UV extinction bump in theoretical studies \citep{li01} and lab experiments \citep{joblin92}.

\begin{figure*}[th]
	\centering
		\includegraphics[width=.4\textwidth,trim={0.1cm 0.2cm 0.1cm 0.2cm},clip]{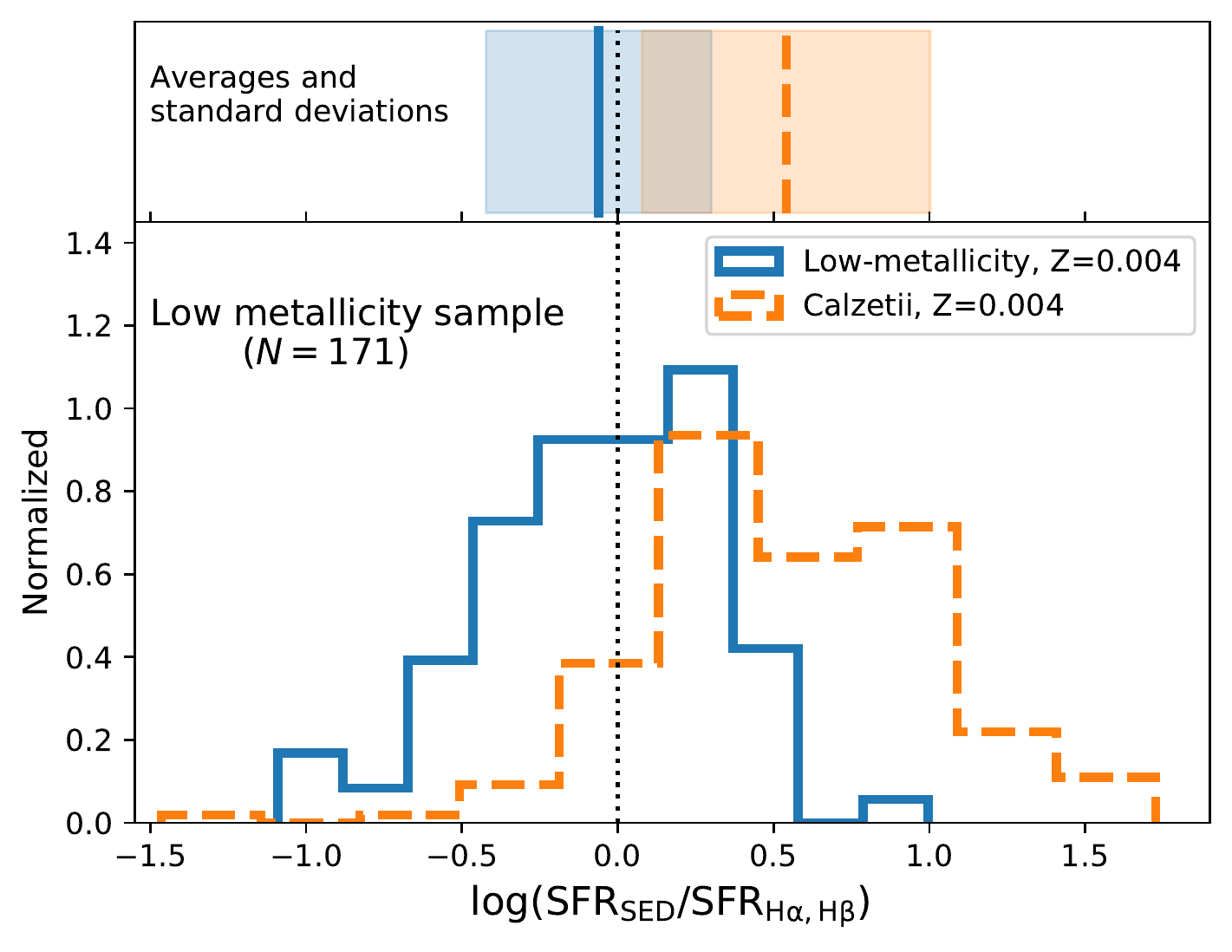}\quad
		\includegraphics[width=.4\textwidth,trim={0.1cm 0.2cm 0.1cm 0.2cm},clip]{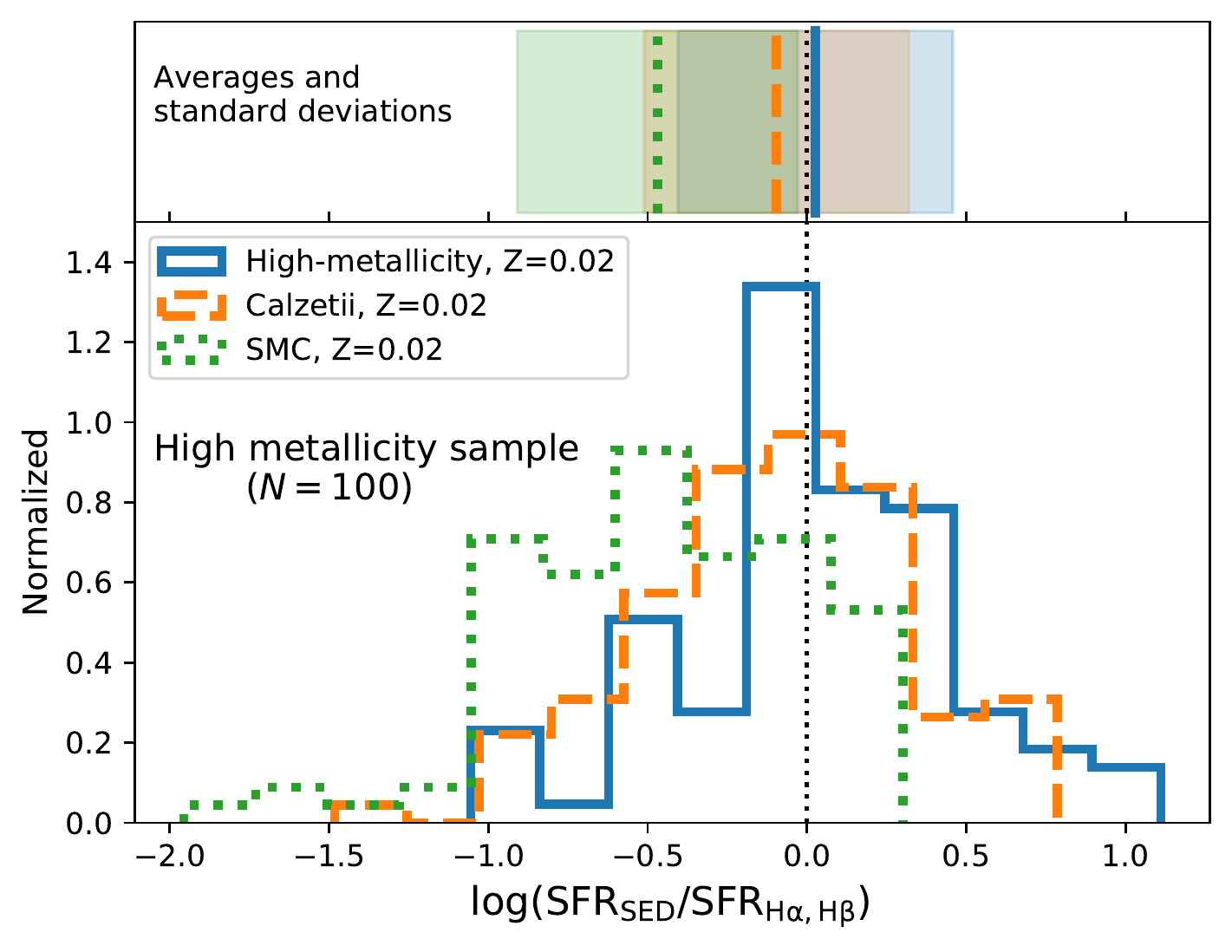}
		\caption{Comparison between the SED-inferred SFRs assuming different attenuation curves with {\sfr} for the low metallicity (left) and high metallicity (right) samples. The upper panels show the average values and standard deviations for each distribution in corresponding colors.
		The SED fitting is performed on galaxies at $z=1.4-2.6$ in CANDELS fields with 3D-{\em HST} photometry and spectroscopic redshifts, {\sfr}, and [N{\sc ii}]/{\halpha} metallicities from the MOSDEF survey.  The number of galaxies in each sample is shown on the plots. We fixed the stellar metallicity in the SED fitting to $Z=0.004$ and 0.02 (solar) for the low- and high-metallicity samples, respectively. The SFRs from the low- and high-metallicity curves agree very well with {\sfr}. The C00 \citep{calzetti00} SFRs are in good agreement with {\sfr} for the high metallicity sample, but slightly overestimate SFRs in the low metallicity sample. A steep SMC curve underestimates SFRs in the high metallicity sample.
		}
		\label{fig:sedcomp}
\end{figure*}

 \subsection{Star Formation Rate Comparison} \label{sec:sfr}

SFR is a key parameter to describe galaxies and their cosmic evolution. The most widely used SFR indicator at high redshifts is UV stellar continuum, as it redshifts to longer wavelengths that are easily accessible from the ground, and can be detected down to lower luminosities given the higher sensitivity of optical-near-IR instruments compared to that of the far-IR observatories. However, the UV is highly affected by dust attenuation and SFRs derived from fitting SED models to the photometry are sensitive to the assumed attenuation curve. In this section, we compare SED-inferred SFRs, assuming our metallicity-dependent curves and the SMC and C00 curves, with independently-derived {\sfr}.

The {\halpha} nebular line is a robust indicator of recent star formation once accurately corrected for attenuation, even for highly star-forming galaxies at $z\sim 2$ \citep[e.g.,][]{muzzin10,reddy10,shivaei15a,shivaei16}. Our MOSDEF sample has 441 non-AGN targets with spectroscopic {\halpha} and {\hbeta} line measurements ($S/N>3\sigma$) at $z=1.4-2.6$. We correct {\halpha} luminosity for attenuation using the Balmer decrement and assuming the \citet{cardelli89} MW curve \citep[see][]{reddy15,shivaei15b,shivaei16}. The MW curve is preferable for the extinction correction of nebular lines, as it is derived based on the line-of-sight measurements of H{\sc ii} regions \citep{calzetti94,wild11b,salim20} and shown to be an accurate representation of nebular attenuation curve at $z\sim 2$ (Reddy et~al., in prep).

We select galaxies that have metallicities from [N{\sc ii}], which results in a sample of 271 galaxies (the additional criteria that were imposed to derive the attenuation curve in Section~\ref{sec:sample} are not used here), and compare the SFRs for the two subsamples of low- and high-metallicity galaxies separated at $12+\log(\rm{O/H})=8.5$ (average $\langle12+\log(\rm{O/H})\rangle$ of 8.4 and 8.6, respectively). To determine SFRs from dust-corrected {\halpha} luminosities, we use $L$({\halpha})-to-SFR conversions derived from BPASSv2.2 SPS models \citep{stanway18}, assuming a Chabrier IMF with upper mass cutoff of 100\,{\msun} and $Z_*=0.02$ and 0.004 for the high- and low-metallicity samples, respectively \citep{theios19}. The solar metallicity ($Z_*=0.02$) conversion is only different by 0.05\,dex from the conversion factors of  \citet{hao11} and \citet{kennicutt12}, once corrected for IMF differences \citep[see Table 2 of][]{theios19}.

To synthesize the unattenuated UV continua, we use the FAST SED fitting code \citep{kriek09a}, with the library of \citet{bc03} stellar population synthesis models, solar ($Z=0.02$) and sub-solar ($Z=0.004$) stellar metallicities for the high- and low-metallicity samples respectively, a \citet{chabrier03} IMF, and delayed exponentially declining star-formation history (SFR $\sim t~{\rm exp(}-t/\tau{\rm )}$, where $t$ is age and $\tau$ is the exponential timescale). The $A_V$ is allowed to vary between 0 and 4 in increments of 0.1. Using the BPASS binary models instead of the \citet{bc03} models would not significantly alter the SED-inferred SFRs, as the effect of binary evolution is much less significant on UV continuum compared to the nebular lines \citep[e.g.,][]{stanway16,shivaei18,theios19}. Figure~\ref{fig:sedcomp} shows the results of the SED-inferred SFRs assuming various attenuation curves compared with the {\sfr}. The SFRs from the low- and high-metallicity curves agree well with the {\sfr} in both of the samples. The C00 curve tends to over-estimate the SFR in the low-metallicity sample, but agrees well with the {\sfr} in the high-metallicity sample. The SMC curve with a solar metallicity stellar population underestimates the SFR in the high-metallicity sample. We do not show the results of the SMC curve with sub-solar stellar metallicity for the low-metallicity sample, as the shape of the low-metallicity curve in the UV is identical to the SMC curve.

In conclusion, this study suggests that on average, a steep attenuation curve, similar to the SMC curve, is more applicable to the low-metallicity (and hence low-mass) high-redshift galaxies, while a shallower C00 curve is preferred for the higher metallicity/mass galaxies. The metallicity division we used in this study is at $12+\log{\rm (O/H)}=8.5$ and corresponds to $M_*\sim 10^{10.4}$\,{\msun}. We caution that the cut in metallicity should not be taken as a clear cut in stellar mass, due to the scatter in the mass-metallicity relation \citep{sanders18}.
We note that SFR calculations are also affected by parameters other than the attenuation curve, such as stellar metallicity, stellar multiplicity, and the shape of the IMF. However, we showed in \citet{shivaei18} that the variations between the {\sfr} and UV-inferred SFRs (which is the same as our SED-inferred SFRs within the uncertainties) are on average dominated by the uncertainties in the adopted attenuation curve, compared to those associated with stellar metallicity, multiplicity, and IMF.

\begin{figure}[tbp]
	\centering
	\subfigure{
		\includegraphics[width=1\columnwidth]{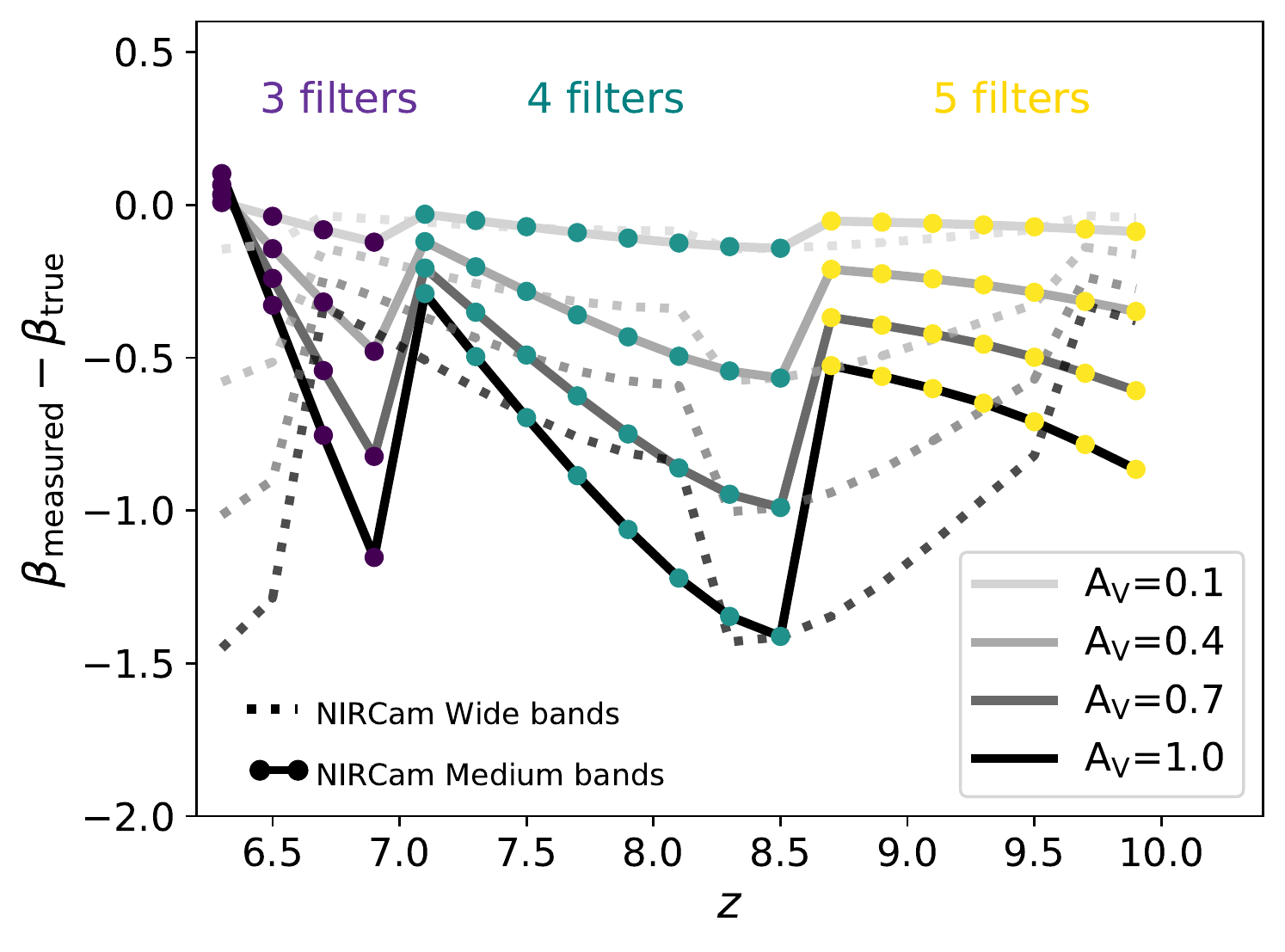}}
	\\
	\subfigure{
		\includegraphics[width=.95\columnwidth]{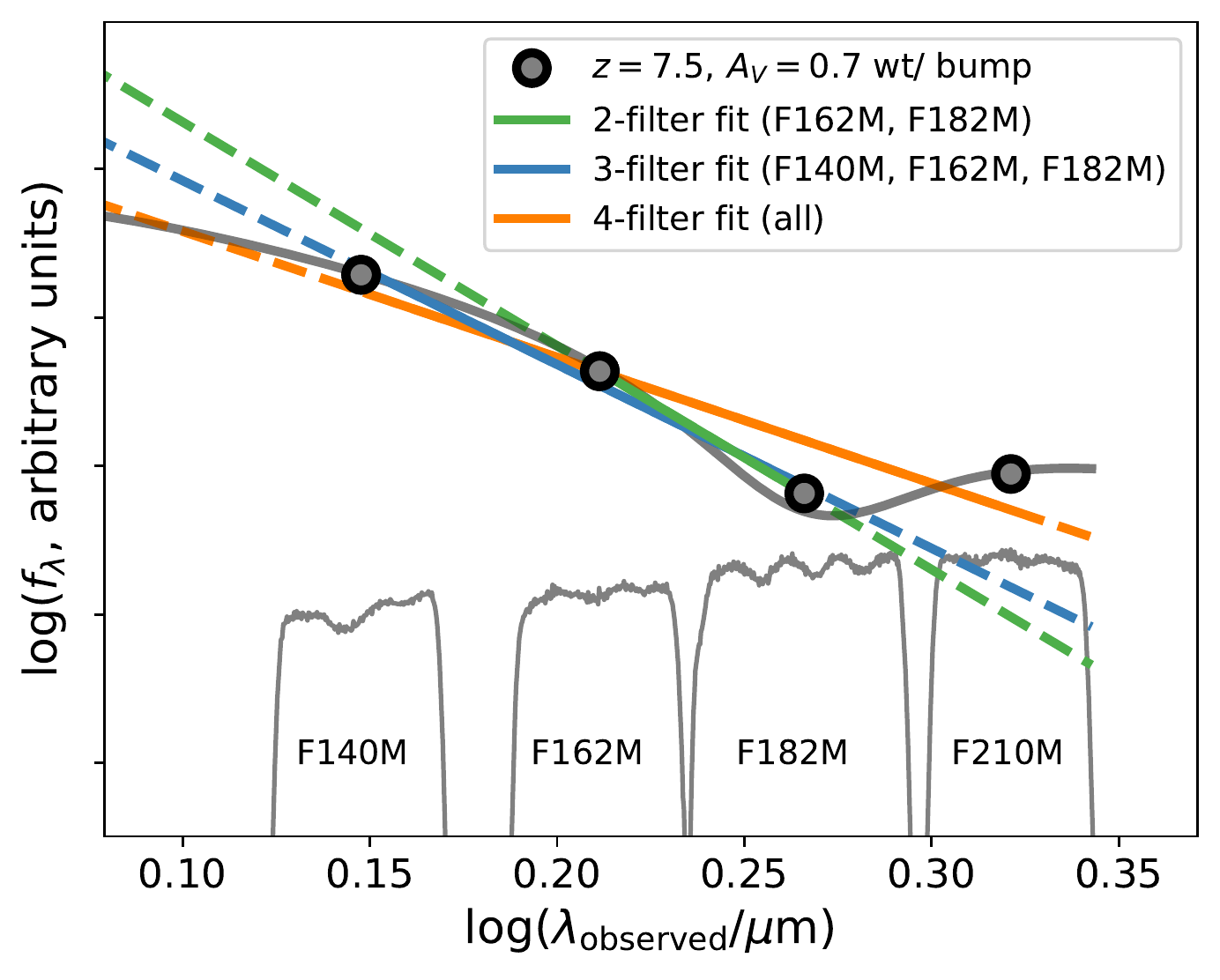}}
		\caption{The effect of the presence of a bump in measuring $\beta$ for high-redshift galaxies with {\em JWST}/NIRCam filters.
		{\em Top:} Difference in the measured and true UV slopes ($\beta$) in the presence of a UV bump similar to the high-metallicity curve bump, as a function of redshift (horizontal axis) and different $A_V$ (lines with shades of grey). $\beta$ is calculated by fitting to the {\em JWST}/NIRCam medium-band fluxes (circles and solid lines) and wide-band fluxes (dashed lines) at rest-frame 1260-2600\,\AA~as $\log(f_{\lambda})\propto \beta \log(\lambda)$. In the case of the medium-band filters, the absence of photometric coverage of the continuum longward of the bump ($\lambda>2400$\,\AA) affects the linear fit to $\log(f_{\lambda})-\log(\lambda)$ and underestimates the UV slope (i.e., smaller $\beta_{\rm measured}$ compared to $\beta_{\rm true}$). The jumps at $z\sim 7.$ and 8.5 in the medium-band filters occur when a filter is added red-ward of the UV bump. This effect is demonstrated in the lower panel. The situation with wide-band filters is similar, but note that the wide-band filters have a poorer sampling of the UV slope window, with only 2 filters at $z=6.3-6.7$ and $8.3-9.7$ and 3 filters in between.
		{\em Bottom:} Difference in the slope ($\beta$) of the linear fit to $\log(f_{\lambda})-\log(\lambda)$ in three cases where the continuum on the both sides of the bump is covered (orange line) or not (blue and green lines). The grey curve is the spectrum of a $z=7.5$ galaxy with intrinsic flux of $f_{\lambda}\propto \lambda^{-2.44}$, attenuated assuming the high-metallicity curve and $A_V=0.7$. The circles are the calculated photometry in {\em JWST}/NIRCam filters shown at the bottom. The lack of photometric coverage of the continuum biases the measured $\beta$ and in this case, resulting in a bluer (steeper) $\beta$. This effect is shown quantitatively in the upper panel. The lack of any coverage blue-ward of the bump would result in a redder (shallower) $\beta$.
		}
		\label{fig:uvslope}
\end{figure}

\subsection{UV Continuum Slope} \label{sec:beta}

The UV continuum slope ($\beta$) is commonly used as a tracer of the attenuation of the stellar continuum, particularly at high redshifts as the rest-frame UV becomes accessible to optical/near-IR instruments. The importance of accurate $\beta$ measurements and the assumed attenuation curve in correcting the observed UV SFRs for dust attenuation has been demonstrated in many previous studies \citep[e.g.,][]{calzetti94,meurer99,reddy06a,boquien12,kriek13,reddy15,shivaei15b,popping17,safarzadeh17,reddy18a,shivaei18,fudamoto19}.
The UV slope is measured by fitting a linear function to $\log(f_{\lambda})$ versus $\log(\lambda)$ at rest-frame $1260-2600$\,\AA, a range that includes the UV 2175\,{\AA} bump.

The presence of a UV bump affects medium-band photometry and the corresponding $\beta$ measurements. 
As an example, we look into the effect of the bump for $z\sim 2$ observations. 
At $z=1.8$, for a set of $G$, $V$, $R$, and $I$ photometric bands, the UV extinction bump falls into the $R$ band, while the $I$ band covers the continuum redward of the bump. Assuming a UV bump similar to that of our high-metallicity curve, the UV slope derived from $G$, $V$, and $R$ bands would be underestimated compared to the UV slope derived from the photometry including the $I$ band by $\Delta \beta=$ 0.2, 0.6, and 0.8 for $A_{V}$ of 0.2, 0.5, and 0.7, respectively. 
With all four bands, it is possible to estimate both the UV slope and the strength of the bump.
This is a similar effect to that demonstrated for {\em JWST}/NIRCam photometry of higher redshift galaxies in Figure~\ref{fig:uvslope}, as explained below.

Although the high-metallicity galaxies (with metallicities of $\sim Z_{\odot}$) are rare at high redshifts, they still exist, for example in the case of high-redshift quasars \citep{jiang07,juarez09} or luminous dusty galaxies \citep{nagao12,debreuck19}. To illustrate this effect, we therefore show an example with the {\em JWST} NIRCam filters for galaxies at $z\sim 6-10$. Depending on the redshift and the available set of photometry, similar results can be obtained on lower redshift galaxies with shorter wavelength medium-band filters. 

We model the intrinsic (dust-free) spectrum of a galaxy to be a power-law function of wavelength with a given intrinsic power ($\beta_0$). The exact choice of $\beta_0$ is irrelevant for our purpose, as we are only interested in the difference between the multiple $\beta$ estimates (hence, $\beta_0$ cancels out in the equations). 
The intrinsic $f_{\lambda}$ is attenuated assuming four $A_V$ values ($A_V=0.1, 0.4, 0.7, 1.0$) and two attenuation curves: 1) the high-metallicity curve with a bump with $\sim 0.5$ the strength of the MW bump, and 2) the same curve without the bump, which is identical to the R15 curve. 
We calculate $\beta$ through the {\em JWST}/NIRCam wide (F115W, F150W, F200W, F277W) and medium band filters (F140M, F162M, F182M, F210M, F250M) filters for targets at $z=6.3-10.0$ with different $A_V$ values, assuming the two attenuation curves with and without a bump (Figure~\ref{fig:uvslope}). 
Similar results are obtained when fitting the rest-frame UV photometry of lower redshift galaxies from {\em HST} and ground-based optical telescopes \citep[e.g.,][]{popping17,tress18}.

The upper panel of Figure~\ref{fig:uvslope} shows the difference in the measured and true $\beta$ as a function of redshift and $A_V$, where $\beta_{\rm measured}$ is calculated from the high-metallicity curve with a bump and $\beta_{\rm true}$ is calculated from the same curve but without the bump.
The measured $\beta$ often underestimates the true $\beta$ (underestimates the true reddening) as the UV 2175\,{\AA} feature moves through the photometric filters. 
The UV slope calculated in the case of the curve with a bump becomes progressively bluer (smaller $\beta$) as the longest-wavelength filter used in the calculations moves into the UV 2175\,{\AA} feature. Once a filter red-ward of the bump is added, the $\beta$ from the curve with a bump gets closer to the true (no-bump) value (notice the jumps at $z\sim 7$ and 8.5 in Figure~\ref{fig:uvslope}-top). This effect is also illustrated in the lower panel of Figure~\ref{fig:uvslope} qualitatively, where a linear function fit to the three photometric points, excluding the F210W filter, results in a bluer $\beta$ compared to the fit to four filters, which captures the continuum on both sides of the UV 2175\,{\AA} feature. A fit to only two bands, where one is affected by the UV 2175\,{\AA} feature, underestimates $\beta$.

As the UV 2175\,{\AA} feature is very broad, if present, it would affect the NIRCam wide-band photometry as well (dotted lines in Figure~\ref{fig:uvslope}). The disadvantage of using the wide-band filters is poorer sampling of the UV continuum window (only 2 or 3 filters at rest-frame $1260-2600$\,\AA), which makes it impractical to mask out the filters that are sensitive to the UV 2175\,{\AA} feature. 

An underestimated (overestimated) $\beta$ at a given $L({\rm IR})/L({\rm UV})$ would resemble a shallower (steeper) attenuation curve in the IRX-$\beta$ relation (relation between $L$(IR)/$L$(UV) and $\beta$) and results in incorrect estimates of dust attenuation. For example, a galaxy that follows the \citet{meurer99} IRX--$\beta$ relation has a $\beta=-1.0$ at ${\rm IRX}=10$. If the measured $\beta$ of this galaxy is underestimated by $\sim 0.5$ ($\beta_{\rm measured}\sim -1.5$) due to the presence of a UV bump similar to the one found in the high-metallicity curve, the observed location of the galaxy will be shifted to the left of the \citet{meurer99} IRX-$\beta$ locus. As a result, this shift may be misinterpreted as bluer $\beta_0$ (for the same attenuation curve) or complex dust-star geometries \citep[e.g.,][]{popping17,reddy18a}.

The relationship between $\beta$ and stellar continuum color excess ($E(B-V)_{\rm stellar}$) assuming the high-metallicity attenuation curve, {\em excluding the bump}, and an intrinsic $\beta$ of $\beta_0=-2.44$ (corresponding to a stellar population with solar metallicity and constant star formation for longer than 100\,Myr) is 
\begin{equation}
\beta = -2.44+4.75\times E(B-V)_{\rm stellar}.
\label{eq:ebv-beta-high}
\end{equation}
The $\beta-E(B-V)_{\rm stellar}$ relationship for the steeper low-metallicity curve and a bluer intrinsic $\beta$ of $\beta_0=-2.62$, which is shown to be more applicable to the low mass/metallicity galaxies at these redshifts \citep{steidel16,reddy18a}, is:
\begin{equation}
\beta = -2.62 + 11.47\times E(B-V)_{\rm stellar}.
\label{eq:ebv-beta-low}
\end{equation}
Different assumptions for the intrinsic $\beta_0$, which depends on the stellar metallicity of the underlying stellar populations, would change the inferred $\beta$ values systematically. Our $\beta_0$ assumptions here are motivated by previous $z\sim 2$ studies in the literature, as mentioned before.

\begin{figure*}[tbp]
	\centering
		\includegraphics[width=.31\textwidth]{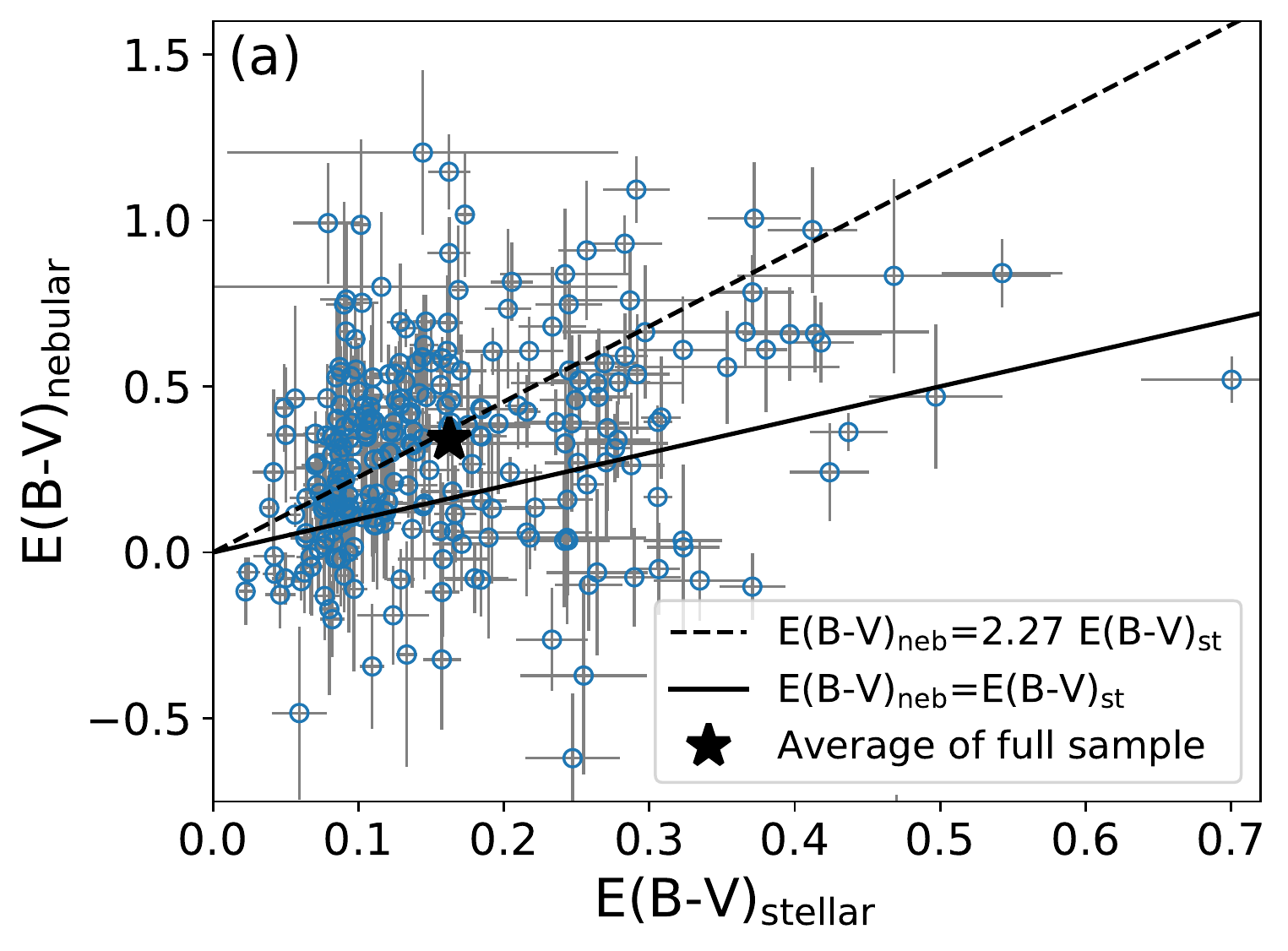}
	\quad
		\includegraphics[width=.31\textwidth]{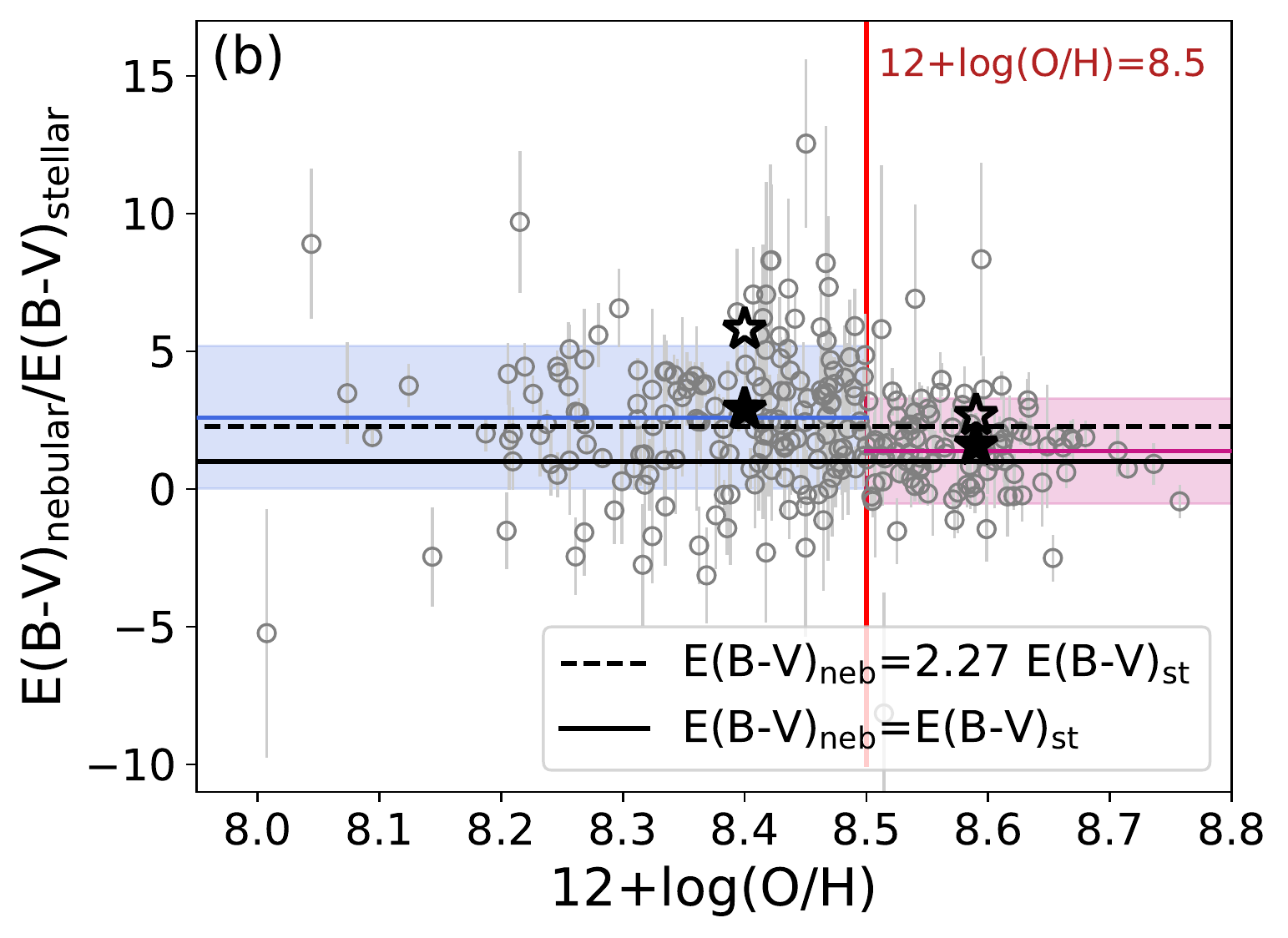}
		\quad
		\includegraphics[width=.31\textwidth]{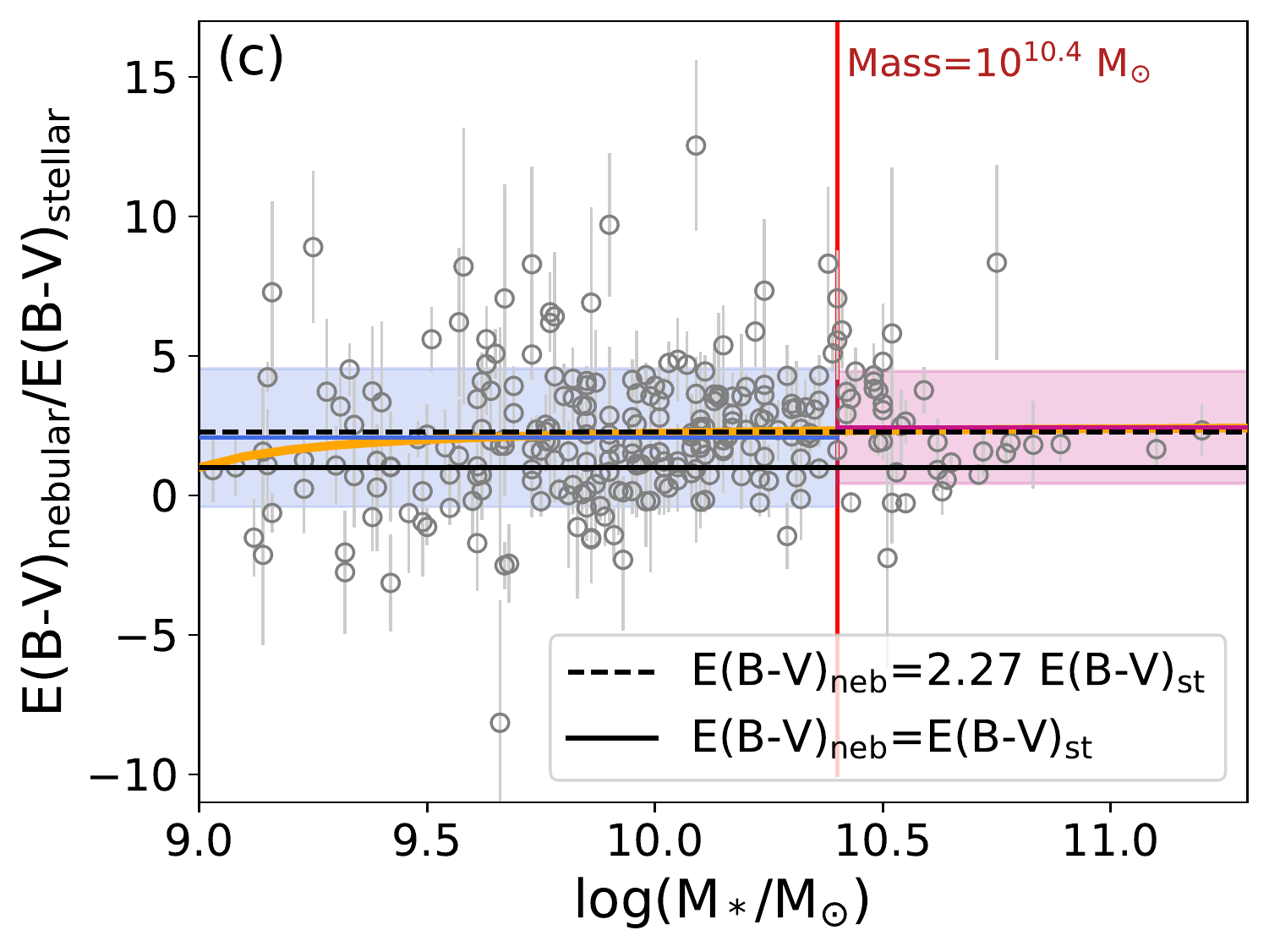}
		\\
		\includegraphics[width=.31\textwidth]{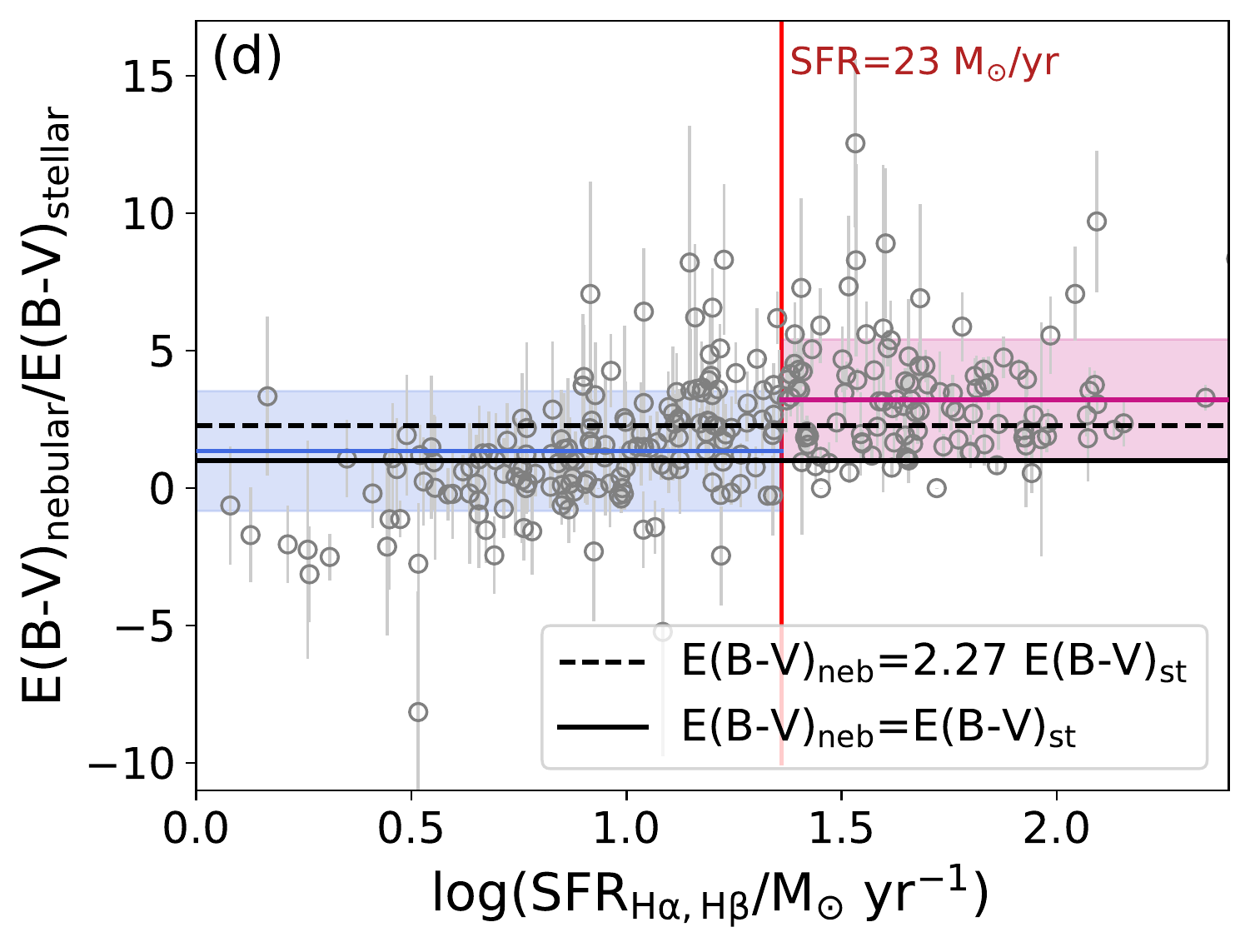}		
		\quad
		\includegraphics[width=.31\textwidth]{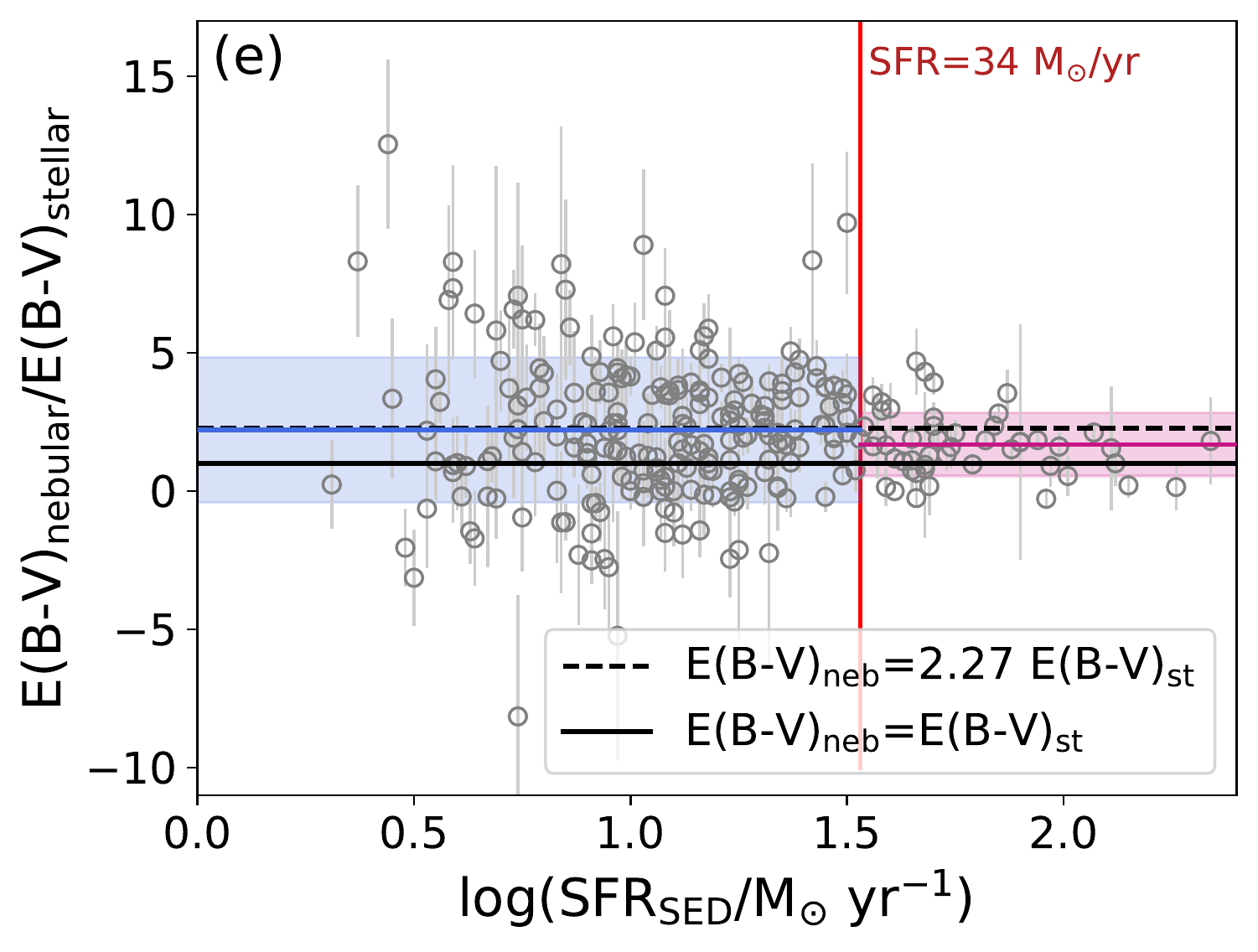}		
		\quad
		\includegraphics[width=.31\textwidth]{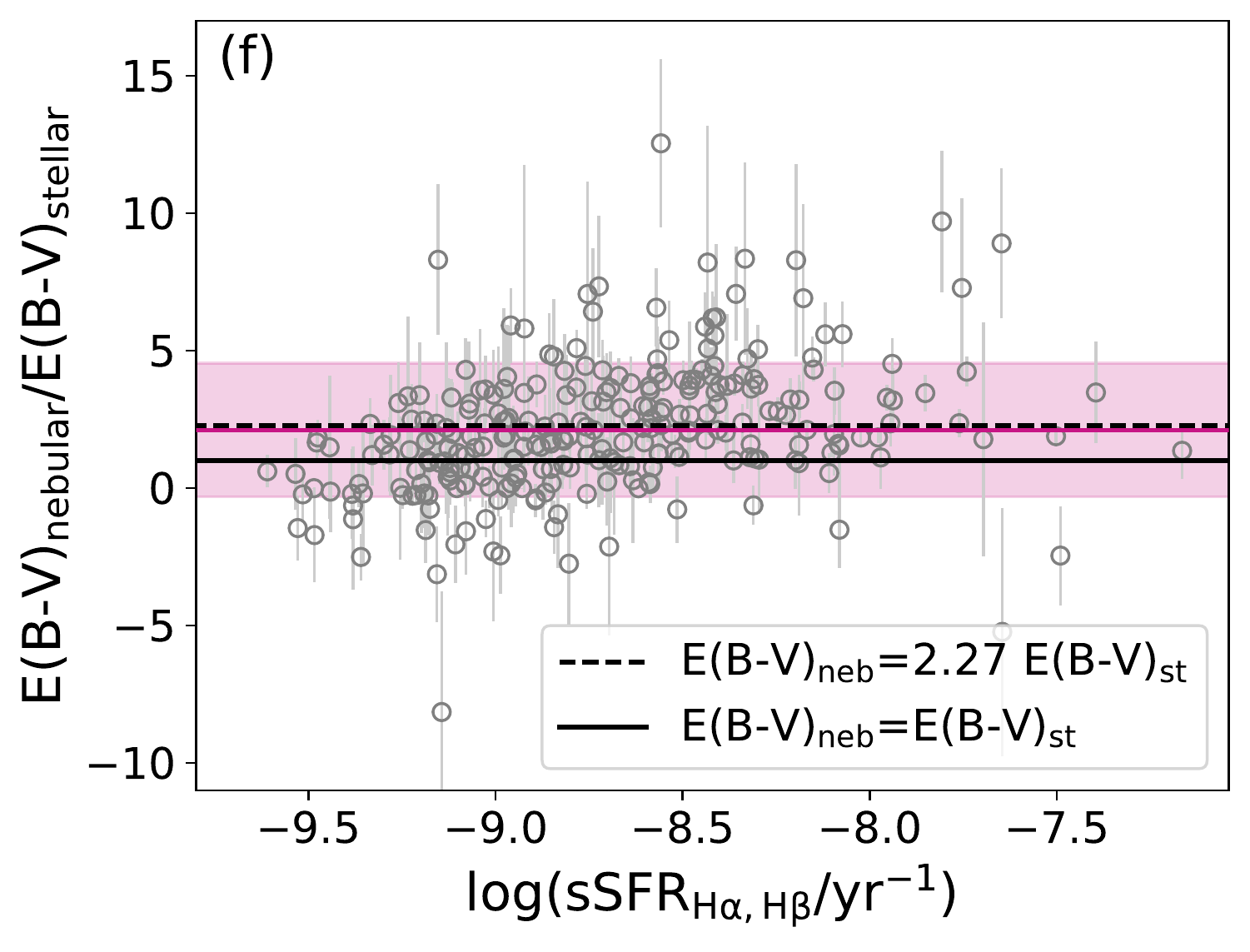}		
		\caption{Comparison of the nebular line and stellar continuum reddening for our sample of 265 galaxies at $z=1.4-2.6$. Nebular reddening is calculated from the Balmer decrement, and the stellar reddening is calculated using Equations~\ref{eq:ebv-beta-high} and \ref{eq:ebv-beta-low} for high and low metallicity galaxies, respectively. {\em (a)}: nebular versus stellar reddening. The linear relation and differential attenuation relation suggested by \citet{calzetti94} are shown with solid and dashed lines. 
		{\em (b-f)}: the ratio of nebular to stellar reddening as a function of gas-phase metallicity (b), stellar mass (c), dust-corrected {\halpha} SFR (d), SED-inferred SFR (e), and sSFR ({\sfr} to mass ratio; f). The average ratios and the standard deviations about the average are shown with horizontal lines and shaded regions. The averages are calculated in two bins in each plot and the division limit is shown with a vertical red line. The metallicity limit is the same as the limit used to derive the attenuation curves in this work. The stellar mass and SFR limits are derived based on MOSDEF mass-metallicity and SFR-mass relations \citep{shivaei15b,sanders18} at the metallicity of $12+\log({\rm O/H})=8.5$. Panel (f) shows the average and standard deviation of the full sample: $E(B-V)_{\rm neb}/E(B-V)_{\rm stellar}=(2.11 \pm 0.15)$ with a standard deviation of 2.43.  The stars in (a) and (b) show the average reddening ratios for the sample of galaxies with undetected {\hbeta} and detected {\halpha} (empty stars) and for the full sample with {\halpha} detection (regardless of the {\hbeta} detection; filled stars), measured from the stacked spectra (see the text). The average reddenings of the full sample ({\halpha}-detected) from the stacked spectrum is 2.12 (($\langle E(B-V)_{\rm neb}\rangle=0.34$ and $\langle E(B-V)_{\rm stellar}\rangle=0.16$), consistent with the average of the {\halpha}-{\hbeta}-detected sample, $2.11\pm 0.15$. The average reddenings in the low- and high-metallicity bins are respectively 2.90 ($\langle E(B-V)_{\rm neb}\rangle=0.29$ and $\langle E(B-V)_{\rm stellar}\rangle=0.10$) and 1.59 ($\langle E(B-V)_{\rm neb}\rangle=0.43$ and $\langle E(B-V)_{\rm stellar}\rangle=0.27$), consistent with the averages of the {\halpha}-{\hbeta}-detected sample (horizontal blue and magenta solid lines in panel b).
		The orange line in (c) shows the ratio of Equation~\ref{eq:dust-mass1} to Equation~\ref{eq:dust-mass2}.
		}
		\label{fig:ebv}
\end{figure*}

\subsection{Nebular versus stellar reddening} \label{sec:ebmv}

Color excess, or reddening, is the difference between the total attenuation at $B$ and $V$ bands, $E(B-V)\equiv{ A_B - A_V}$. In a simple star-dust geometry of a screen of dust in front of the stars, $E(B-V)$ represents the thickness of the dust screen (i.e., the dust column density). The color excess measured from the continuum emission (e.g., using UV continuum slope $\beta$ or SED fitting to the photometry) shows the reddening of the stellar continuum, while the color excess derived from the nebular emission lines indicates the nebular (gas) reddening. If all of the stars in a galaxy are subject to the same dust column densities, the stellar and nebular reddenings should be similar \citep{erb06b,reddy10,pannella15,shivaei15a,pulgisi16}. However, multiple studies have shown a higher nebular reddening compared to the stellar reddening in star-forming galaxies \citep{calzetti94,calzetti00,garn10,mancini11,wild11,kashino13,kreckel13,price14,reddy15,battisti16,buat18,qin19,theios19}. We compile a number of these determinations in Table~\ref{tab:ebvs}. The simplest physical interpretation of a differential attenuation is a two component dust-star model, where all stars experience a modest attenuation due to the diffuse ISM dust, but the young and massive stars that give rise to the nebular lines are embedded in their dense and dusty birthclouds \citep[e.g.,][]{calzetti94,charlotfall00}. 

\capstartfalse   
\begin{deluxetable}{ccc}
    \tablewidth{1.0\hsize}
    \tablecolumns{3}
    \tablecaption{Estimates of $E(B-V)_{\rm neb}/E(B-V)_{\rm stellar}$
    }
    \tablehead{
	\colhead{$E(B-V)_{\rm neb}$/} & Redshift  & Reference \\
   \colhead{$E(B-V)_{\rm stellar}$}  &    &  (chronological order)   
	    }
    \startdata
$\sim$ 2  &  $<$ 0.03  & \citet{calzetti94}  \\
 2.27 $\pm$ 0.16  &   $<$ 0.03   & \citet{calzetti00}  \\
$\sim$ 2 &   $\sim$ 0.84    & \citet{garn10}  \\
$\sim$ 2.2   &  1.4 $-$ 2.5       &  \citet{mancini11}  \\
 2.1 $-$ 2.9  &   $\sim$ 0.07     &  \citet{wild11}  \\
1.20 $\pm$ 0.15  &  1.4 $-$ 1.7  & \citet{kashino13}  \\
1.86$^{+0.40}_{-0.37}$  &   1.5 $-$ 2.5  & \citet{price14}  \\
$\sim$ 1.3  &   $\sim$ 1  &  \citet{pannella15}  \\
1.08 $\pm$ 0.07  &  $\sim$ 1  &  \citet{pulgisi16}  \\
$\sim$ 2  &  $\le$ 0.1  &  \citet{battisti16}  \\
1.85$^{+1.00}_{-0.26}$  &  0.6  $-$ 1.6  &   \citet{buat18}  \\
$\sim$ 1.89  & 0.04 $-$ 0.15  &  \citet{qin19}  \\
$\sim$ 1.34   & 2.0 $-$ 2.7    &  \citet{theios19}  \\
2.11 $\pm$ 0.15  &  1.4 $-$ 2.6   &  This work
    \enddata
\label{tab:ebvs}   
\end{deluxetable}
\capstarttrue  

Figure~\ref{fig:ebv}(a) shows the relation between the two reddenings in our sample with {\halpha}, {\hbeta}, and [N{\sc ii}] detections (271 galaxies, similar to the sample in Section~\ref{sec:sfr}). The reddening of the stellar continuum is derived using the UV continuum slope and Equations \ref{eq:ebv-beta-high} and \ref{eq:ebv-beta-low} for the high- and low-metallicity galaxies, respectively\footnote{The stellar reddening derived from UV slope and Equations \ref{eq:ebv-beta-high} and \ref{eq:ebv-beta-low} closely follows the reddening derived from SED model fitting, using the same attenuation curves.}. The nebular reddening is calculated from the Balmer decrement and assuming a MW curve (Reddy et~al., in prep), as $E(B-V)_{\rm neb}=2.33\,\log(\frac{\rm H\alpha/H\beta}{2.86})$. 
There is substantial scatter in $E(B-V)_{\rm neb}$ that results in a few of the nebular reddening determinations to be negative due to the statistical uncertainties in the {\halpha} and {\hbeta} flux measurements. However, as expected from statistical noise, the negative nebular reddenings are consistent with zero at 3$\sigma$ level. To avoid biasing the data and the statistical inferences, we do not set the negative $E(B-V)$'s to zero\footnote{Our main results will be unaffected by setting the cases with $E(B-V)<0$ to zero (as has been done in some studies). This condition will systematically increase the average $E(B-V)_{\rm neb}$ values, as expected, but does not change any of the main conclusions, as the negative $E(B-V)_{\rm neb}$ values are distributed over all metallicities and masses and are not systematically clustered at a certain metallicity or mass. By setting the negative nebular reddenings to zero, the average reddening ratios ($\langle E(B-V)_{\rm neb}/E(B-V)_{\rm stellar}\rangle$) for the low- and high-metallicity bins will be $2.81\pm 0.17$ and $1.58\pm 0.15$, respectively, and for the low- and high-mass bins will be $2.30\pm 0.14$ and $2.52\pm 0.31$, respectively.}. As seen in Figure~\ref{fig:ebv}(a), there are no corresponding negative values in the stellar reddenings, as in our sample the measured $\beta$ is always larger than the assumed intrinsic $\beta_0$ in Equations \ref{eq:ebv-beta-high} and \ref{eq:ebv-beta-low}. 

The Pearson correlation coefficient in Figure~\ref{fig:ebv}(a) indicates a correlation between the nebular and stellar reddening in our sample with $\rho=0.2$ and a p-value of $\sim 10^{-4}$. The differential attenuation relation of $E(B-V)_{\rm neb}= 2.27\,E(B-V)_{\rm stellar}$, suggested by \citet{calzetti00}, and the unity relation are also shown in Figure~\ref{fig:ebv}(a). The large scatter in the relation (standard deviation of 2.43 from the mean) makes it difficult to favor one relation over the other; however the majority of the galaxies show $E(B-V)_{\rm neb}>E(B-V)_{\rm stellar}$, with an average of $E(B-V)_{\rm neb}=(2.11 \pm 0.15)\,E(B-V)_{\rm stellar}$ for the full sample. 

The derived nebular reddenings do not significantly change under a different assumption of the dust curve, as $E(B-V)_{\rm neb}$ depends on the shape of the dust curve at {\hbeta} and {\halpha} wavelengths ($E(B-V)_{\rm neb}=\frac{2.5}{k_{\rm H\beta}-k_{\rm H\alpha}}\,\log(\frac{\rm H\alpha/H\beta}{2.86})$) and the shapes of various curves, such as MW, LMC, SMC, and C00, at these wavelengths are very similar to each other. The nebular reddening, as mentioned before, is more affected by measurement uncertainties and statistical noise. On the other hand, the derived stellar reddening depends strongly on the assumption of the attenuation curve and the intrinsic $\beta_0$. Smaller stellar $E(B-V)$'s are obtained if an SMC curve is assumed for high-metallicity galaxies \citep[e.g., as assumed in Figure 9 of][]{theios19}. In this study, we have found that a combination of two different attenuation curves is more applicable to galaxies over a wide range of metallicities, and hence, we use the equations \ref{eq:ebv-beta-high} and \ref{eq:ebv-beta-low} to derive the stellar reddenings for $12+\log({\rm O/H})>8.5$ and $<8.5$, respectively. 

We further investigate the ratio of the nebular to stellar reddening (the reddening ratio) as a function of galaxy parameters in Figure~\ref{fig:ebv}. 
We calculate the average and standard deviations of the reddening ratio in different ranges of gas-phase metallicity, mass, and SFR, shown with horizontal lines and shaded regions. In gas-phase metallicity, we divide the sample at $12+\log({\rm O/H})=8.5$, the same metallicity limit that was used in deriving the two attenuation curves.
The stellar mass and SFR limits are calculated based on the MOSDEF mass-metallicity relation \citep{sanders18} and MOSDEF SFR-mass relations for {\halpha} and SED SFRs \citep{shivaei15b} at the metallicity of $12+\log({\rm O/H})=8.5$.

As shown in Figure~\ref{fig:ebv}(d), the reddening ratio correlates with {\sfr} ($\rho=0.6$, p-value$<10^{-7}$). However, the positive correlation may be due to the degeneracy between the two parameters, as $E(B-V)_{\rm neb}$ is used to correct {\sfr} for dust attenuation. The correlation disappears when SED SFR is used instead of {\sfr} (panel e), and instead, the reddening ratio shows a large scatter at low SED SFRs. The SED SFR is also indirectly related to the reddening ratio, as the $E(B-V)_{\rm stellar}$ correlates tightly with the SED-inferred $E(B-V)$, which determines the dust correction in SED SFR. The modest correlation between the reddening ratio and sSFR ({\sfr}/$M_*$, panel f) may also be a consequence of the degeneracy between $E(B-V)_{\rm neb}$ and {\sfr}. Due to the aforementioned degeneracies, and the inconsistent results based on {\halpha} and SED SFRs, we do not draw robust conclusions regarding the correlation between the reddening ratio and SFR or sSFR.  Disentangling the relation between the reddening ratio and SFR requires an independent SFR indicator, such as IR.
The average reddening ratio, shown with the magenta line in panel f, is $E(B-V)_{\rm neb}=(2.11 \pm 0.15)\,E(B-V)_{\rm stellar}$, with a large scatter shown with the magenta region (standard deviation of 2.43). The average of the reddening differences can also be expressed as $\langle E(B-V)_{\rm neb}-E(B-V)_{\rm stellar}\rangle = 0.18 \pm 0.02$. Keeping this conversion in mind, our Figures~\ref{fig:ebv} (a) and (d) agree well with Figures 16 and 17 of \citet{reddy15}.

\begin{figure*}[tbp]
	\centering
		\includegraphics[width=.45\textwidth]{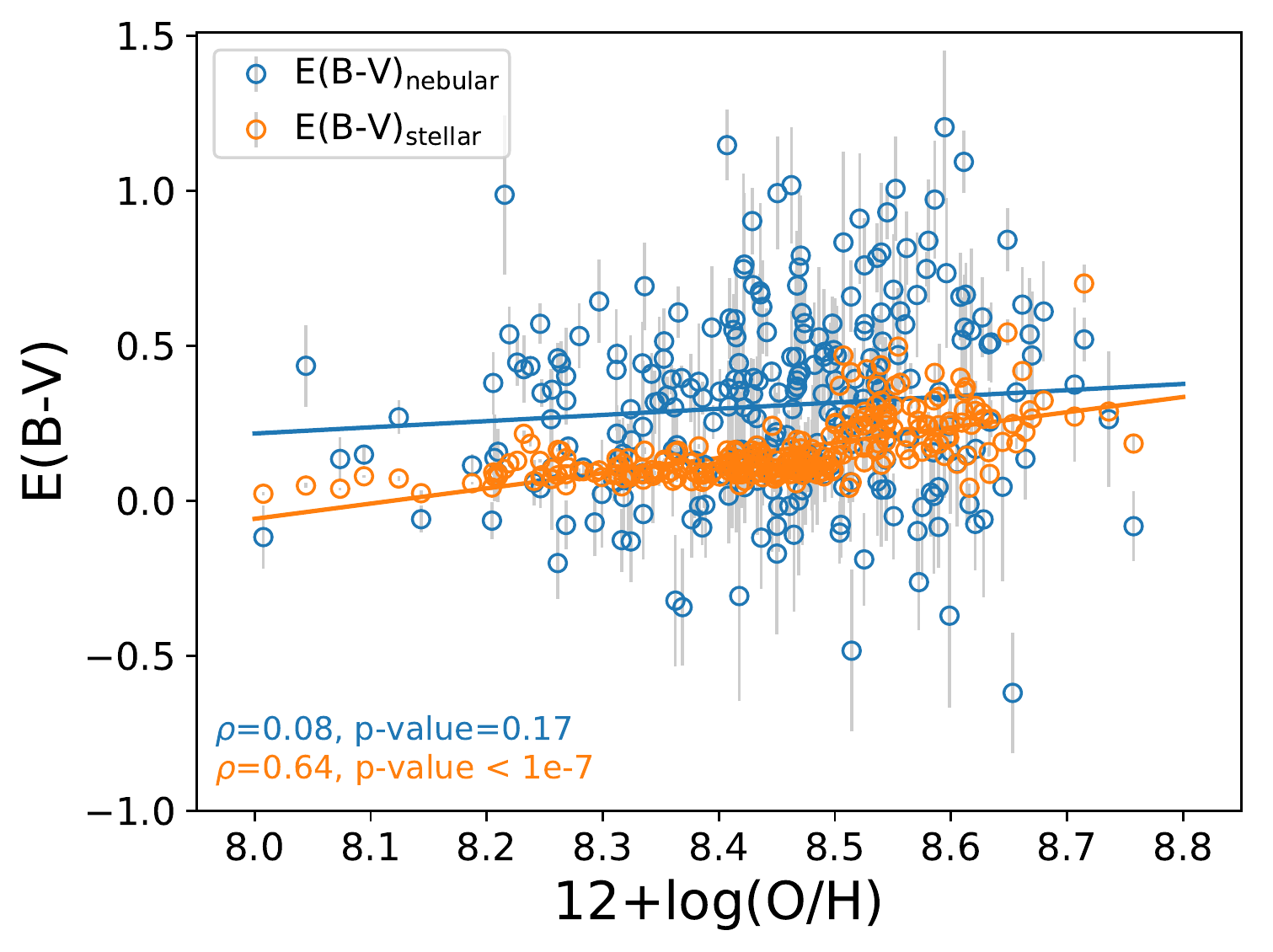}
		\quad
		\includegraphics[width=.45\textwidth]{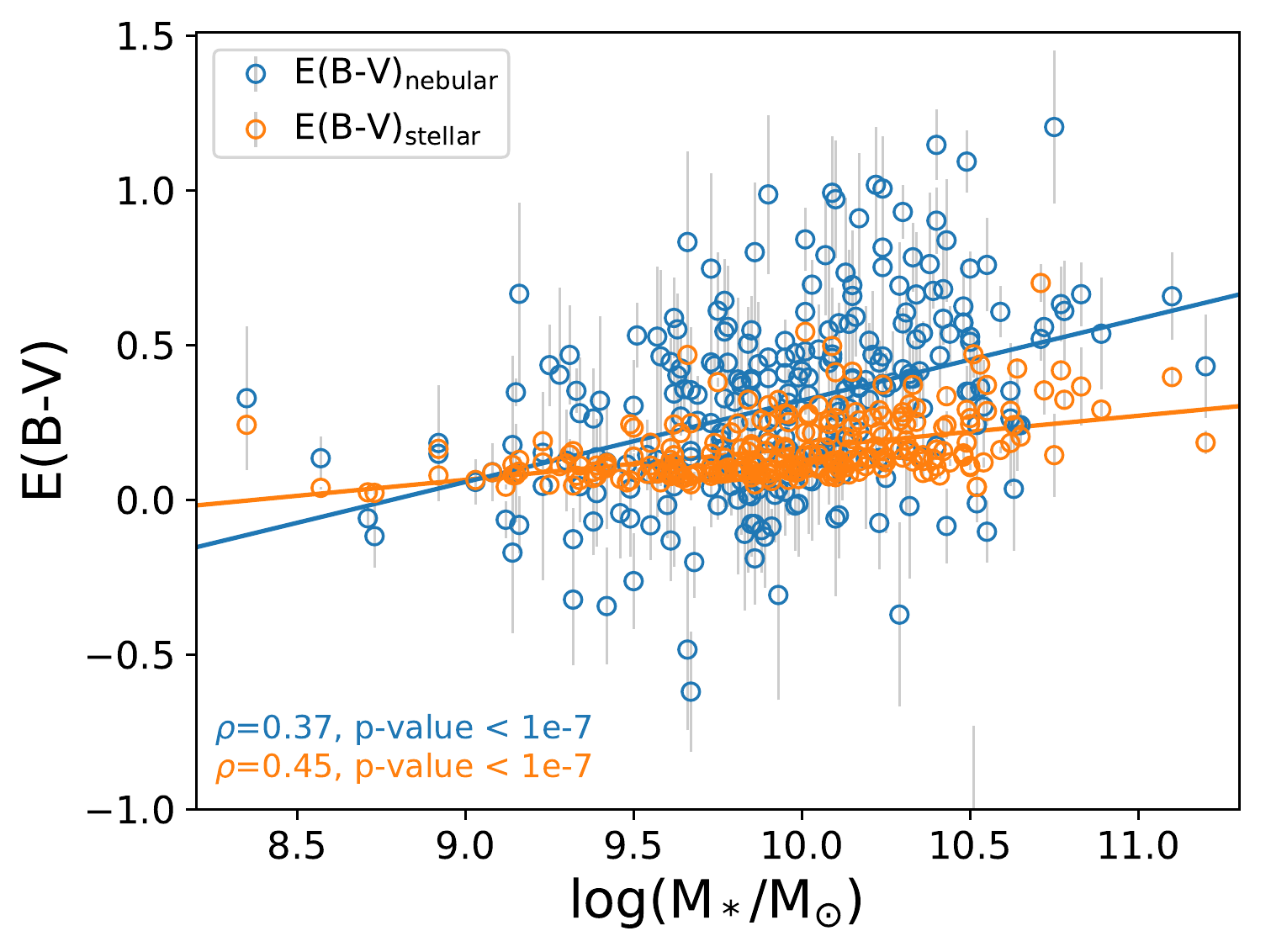}
		\caption{Nebular (blue) and stellar (orange) reddening as a function of metallicity (left) and stellar mass (right). The Pearson correlation coefficient and its p-value for each sample are shown in the bottom-left corner of the plots in the corresponding colors. The linear regression fits are shown with solid lines. Both nebular and stellar reddening increase with increasing metallicity and mass, although the trend is weak for nebular reddening vs. metallicity. The stellar reddening trend is significantly tighter at low metallicities compared to the nebular reddening. The large scatter in nebular reddening may be due to large galaxy-to-galaxy variations in the dust-star geometry and the fraction of very dusty regions around massive stars, while the stellar emission is mainly affected by the ISM diffuse dust component, particularly at low metallicities. 
		}
		\label{fig:ebv-metal}
\end{figure*}

Mass and metallicity are derived independently from the two reddenings. 
Stellar mass is inferred from the best-fit SED model (Section~\ref{sec:sed-sfr}), and is determined by the normalization of the SED model relative to the photometry. The gas-phase metallicity is derived from the ratio of [N{\sc ii}] to {\halpha}. Although metallicity has the uncorrected {\halpha} line luminosity in common with the nebular reddening, the two parameters are not strongly degenerate.

The trend of the reddening ratio with metallicity is interesting (Figure~\ref{fig:ebv}, b). Objects with high metallicity tend to have a similar reddening for line and continuum photons, $E(B-V)_{\rm neb}/E(B-V)_{\rm stellar}=1.38\pm 0.19$\footnote{This value results from using the appropriate attenuation curve, i.e., C00, for the high metallicity sample; use of the SMC curve would raise this value to correspond more closely with the low metallicity part of our sample.} with a standard deviation of $\sigma=1.91$. In comparison, the low-metallicity galaxies show an increase in the reddening ratio, $E(B-V)_{\rm neb}/E(B-V)_{\rm stellar}=2.60\pm 0.20$, and an increased scatter of $\sigma=2.59$. 
This trend is consistent with the findings of \citet{kashino13}, \citet{pannella15}, and \citet{pulgisi16}, who found similar nebular and stellar reddenings, based on samples in which a considerable majority of galaxies have $\log(M_*/M_{\odot})>10.4$. As discussed in Section~\ref{sec:total}, these galaxies would fall within our high-metallicity bin. Most of the other studies cited earlier are extended to significantly lower masses \citep{garn10,mancini11,wild11,price14,battisti16,qin19}, thus accounting for the larger values they obtain for $E(B-V)_{\rm neb}/E(B-V)_{\rm stellar}$ ($\sim 1.9-2.9$), consistent with our average high value of $E(B-V)_{\rm neb}/E(B-V)_{\rm stellar}=2.11$ over the full range of metallicities ($12+\log({\rm O/H})\sim 8.0-8.8$). 

We do not see significant correlation between the reddening ratio and stellar mass (Figure~\ref{fig:ebv}, c), in agreement with \citet{reddy15}, Figure 17. The average reddening ratio is $2.07\pm 0.17$ and $2.44\pm 0.34$ with a scatter of 2.48 and 2.01 respectively for the masses below and above $\log(M_*/M_{\odot})=10.4$, corresponding to the dividing line between the low- and high-metallicity bins (Section~\ref{sec:total}). We suspect that the lack of correlation between the reddening ratio and mass, despite the correlation between the reddening ratio and metallicity, is due to: (1) the relatively small number of galaxies in our sample with $\log(M_*/M_{\odot})>10.4$; and (2) the large scatter in the mass-metallicity relation, particularly when [N{\sc ii}]/{\halpha} is adopted as the metallicity diagnostic \citep{sanders18}. 

Further insight to the trend between reddening and metallicity is provided in Figure~\ref{fig:ebv-metal}, left. A positive correlation between dust reddening and metal abundance is expected, due to the increase of dust to gas mass ratios, as well as the increase of molecular gas fractions, with oxygen abundances \citep{draine07b,leroy11}. 
However, contrary to expectation, the correlation between the nebular reddening and metallicity in our sample is weak (Figure~\ref{fig:ebv-metal}; Pearson $\rho = 0.08$ with a p-value of 0.17), while the stellar reddening shows a significantly stronger correlation with metallicity ($\rho = 0.68$, p-value $< 10^{-7}$). Additionally, at $12+\log({\rm O/H})<8.5$, the stellar reddening is lower with a very small scatter ($\langle E(B-V)_{\rm stellar} \rangle=0.12, \sigma=0.04$) compared to the nebular reddening ($\langle E(B-V)_{\rm neb} \rangle=0.29, \sigma=0.27$), while at higher metallicities the average and scatter of stellar reddening increase and approach to those of the nebular reddening ($\langle E(B-V)_{\rm stellar} \rangle=0.26, \sigma=0.11$ and $\langle E(B-V)_{\rm neb} \rangle=0.33, \sigma=0.38$). 
The reduced scatter of the stellar reddening at low metallicities might be partly affected by the choice of a steeper attenuation curve, which tends to result in lower $\beta$-inferred color excesses (Equation~\ref{eq:ebv-beta-low}). However, we also see an increase in the scatter of the observed $\beta$ with metallicity in our sample. Further investigation on this correlation is beyond the focus of this paper and will be discussed elsewhere (Shapley et~al., in prep).

\citet{pannella15} suggested that the close to unity ratio of $E(B-V)_{\rm neb}/E(B-V)_{\rm stellar}$ in their high-redshift sample ($z\gtrsim 1.5$, similar behavior seen in \citealt{erb06b}, \citealt{reddy10}, and \citealt{shivaei15a}) is a result of high SFR surface densities (higher sSFRs and smaller sizes) of galaxies at $z>1.5$ compared to local galaxies \citep[e.g.,][]{whitaker12b,vanderwel14,rujopakarn16,elbaz18}. As mentioned before, the high-redshift sample of \citet{pannella15} has masses of $M_*\gtrsim 10^{10.5}$\,{\msun}, consistent with the masses of our high-metallicity sample. In such systems, the high star formation surface density and larger dust-to-gas ratios result in a dense ISM that evidently reddens the stellar continuum photons to a similar degree as the reddening of the nebular photons that emerge from the dense birth clouds of newly-born stars. This results in divergence from the two component dust-star model and hence, $E(B-V)_{\rm neb}\sim E(B-V)_{\rm stellar}$. Such an effect is weaker at lower metallicities, due to both the reduced dust-to-gas mass ratios \citep{remyruyer14} and lower scattering and absorption coefficients of different dust grains that are present in low-metallicity galaxies compared to the high-metallicity ones \citep{weingartner01a}. Hence, at low metallicities, the continuum photons are mainly reddened by the uniformly-distributed diffuse dust in the ISM, which is nearly transparent in these galaxies and results in low stellar reddenings. Meanwhile, the line photons, originating from the H{\sc ii} regions, are still heavily obscured as high gas densities are still required for massive star formation, therefore resulting in $E(B-V)_{\rm neb}>E(B-V)_{\rm stellar}$ at low metallicities. This situation is illustrated by extinction maps of the SMC \citep{dobashi09}, which show a few localized spots of very high extinction in what is otherwise a very low level extinction across the galaxy. The two-component model is preserved in this case because of the finite lifetimes of these birth clouds \citep{charlotfall00} and the low ISM absorption once they dissipate.

In this physical picture, the low opacity of the ISM in low-metallicity galaxies outside a small number of actively star-forming regions is consistent with the larger scatter in the nebular-to-stellar reddening ratio at low metallicities ($\sigma=2.59$) compared to that at high metallicities ($\sigma=1.90$, shaded regions in Figure~\ref{fig:ebv}b). A similar trend of decreasing the scatter of the reddening ratio is seen with increasing stellar mass and SED SFR in Figure~\ref{fig:ebv}. The large intrinsic scatter in the attenuation curve at low metallicities discussed in Section~\ref{sec:slope-unc} is also related to the large scatter at low metallicities in Figure~\ref{fig:ebv}(b), both suggesting large galaxy-to-galaxy variations in dust-star geometry at low metallicities with potentially younger ages (see also the discussion in Section~\ref{sec:imp-dust}).

A similar trend between the reddenings and metallicity was seen in the $z=2.0-2.6$ sample of \citet{theios19}. These authors attributed the trend to the line ratios, such as [N{\sc ii}]/{\halpha}, being more sensitive to the stellar abundance of Fe, and not the gas-phase oxygen abundance \citep{steidel14,sanders16a,strom17}. In that scenario, the tight correlation between the stellar reddening and the [N{\sc ii}]/{\halpha} ratio is explained by both quantities being closely tied to the spectral shape of the ionizing radiation field produced by the massive stars.

Nebular reddening has a tighter correlation with stellar mass compared to metallicity, with a Pearson $\rho=0.37$ and a high significance (p-value $\sim 10^{-10}$). As a useful empirical relation \citep[as shown in, e.g.,][]{garnbest10,whitaker17,mclure18}, we fit the reddening versus stellar mass data with a linear function:
\begin{eqnarray}
E(B-V)_{\rm neb}=0.26\,\log(M_*/M_{\odot})-2.31, \label{eq:dust-mass1} \\
E(B-V)_{\rm stellar}=0.10\,\log(M_*/M_{\odot})-0.87.
\label{eq:dust-mass2}
\end{eqnarray}
Caution should be taken when using these equations, given the large scatters. The scatter from the best-fit line in the nebular reddening-mass equation is 0.28 and the scatter in the stellar reddening fit is 0.09. Figure~\ref{fig:ebv-metal}-right confirms that the reddening, and hence dust content of galaxies, increases with mass, for both the stellar continuum and the nebular emission. The ratio of the two fits (Equations~\ref{eq:dust-mass1} to \ref{eq:dust-mass2}, i.e., nebular to stellar reddening as a function of mass) is shown in Figure~\ref{fig:ebv}c (orange curve).

In summary, calculating the continuum reddening is highly sensitive to the assumed attenuation curve. We exploit our metallicity-dependent attenuation curves to derive the continuum reddening and compare it to the nebular reddening calculated from Balmer decrement. On average, we observe a differential attenuation, where the line photons are $\sim 2$ times more reddened than the continuum photons. However, the differential attenuation is less significant in the high metallicity sample compared to low metallicity galaxies, which is likely due to the higher surface density of very dusty regions that redden both the continuum and nebular photons in high-metallicity galaxies. At low metallicities, where galaxies are also preferentially younger, the enhanced differential attenuation suggests a two component dust-star model with older stars reddened by the transparent ISM dust and younger stars are embedded in their dust cocoons.
In any case, the reddening shows a large scatter at a given mass or metallicity, indicating a large galaxy-to-galaxy variation between the nebular and stellar reddenings.

Using data available at the time, \citet{reddy15} found an increasing difference in nebular and stellar reddening with increasing SFR and dust attenuation, similar to the trends we see in Figure~\ref{fig:ebv} with {\sfr} and SFR(SED). Our new analysis suggests that the ratio of the nebular and stellar reddening is smaller at higher gas-phase metallicities (and hence, higher dust attenuation). This difference in interpretation stems mainly from a direct analysis of $E(B-V)_{\rm neb}/E(B-V)_{\rm stellar}$ vs. gas-phase metallicity, and our use of a gas-phase-metallicity-dependent attenuation curve. Direct observations of dust obscuration, such as from the ratio of IR-to-UV luminosity, should help to clarify the how $E(B-V)_{\rm neb}/E(B-V)_{\rm stellar}$ is connected to dust content, gas-phase metallicity, stellar mass, and SFR.

We note that our conclusions, by definition, may not hold for IR-selected samples with heavily obscured objects, such as those in the majority of ALMA extragalactic surveys. Although for this part of the analysis we do not impose the criteria we used for the curve derivation (Section~\ref{sec:sample}) and we include galaxies with {\taub} $>1$, the sample is still selected based on the detection of optical emission lines and is biased against objects that are heavily obscured. In panel (b) of Figure~\ref{fig:ebv} we show the reddening ratio of objects with undetected ($<3\sigma$) {\hbeta} lines ({\halpha} is detected) and the full sample of detected-{\halpha} galaxies (regardless of detection in {\hbeta}) with empty and filled stars, respectively. To derive the Balmer decrement for these galaxies, we normalize the individual spectra to their {\halpha} luminosities, stack the normalized spectra, and measure the {\hbeta} luminosity by fitting a Gaussian function. The inverse of the normalized stacked {\hbeta} luminosity is taken as the average $\langle${\halpha}/{\hbeta}$\rangle$ ratio. For details of the stacking technique refer to Section 3.1 of \citet{shivaei18}\footnote{Source code available at \url{https://github.com/IreneShivaei/specline}}. The reddening ratio inferred from the stacked spectrum of the {\hbeta}-undetected sample (empty star in Figure~\ref{fig:ebv}-b) is larger than the average reddening ratio of the {\hbeta}-detected sample (solid horizontal line in Figure~\ref{fig:ebv}-b), indicating an increased nebular attenuation, but the trend with metallicity is not changed (Figure~\ref{fig:ebv}-b). The average reddenings of the full sample of {\halpha}-detected galaxies (filled stars in Figure~\ref{fig:ebv}-b) are 2.90 and 1.59 for the low- and high-metallicity bins, respectively, consistent with the averages of the {\halpha}- and {\hbeta}-detected sample (horizontal lines; 2.60$\pm 0.20$ and 1.38$\pm 0.19$ for the low and high metallicities, respectively). The majority of our {\hbeta}-undetected galaxies have sky line contamination in their {\hbeta} lines, which may explain the low signal-to-noise. Further investigations of the reddenings of galaxies with large optical depths require IR-selected samples. 

\section{Summary} 
\label{sec:summary}

In this work, we present constraints on the shape of the UV-optical stellar attenuation curve as a function of gas-phase metallicity at $z\sim 2$. Our sample consists of 218 star-forming galaxies at $z=1.4-2.6$ with robust {\halpha}, {\hbeta}, and [N{\sc ii}] spectroscopic line measurements from the MOSDEF survey. The sample is carefully selected to exclude galaxies that show AGN activity, the quiescent populations, ULIRGs and heavily-obscured galaxies, and those with high emission line equivalent widths and young stellar populations.
Following the method of \citet{calzetti00} and \citet{reddy15}, we derived the effective attenuation curve (i.e., $k_{\lambda}-R_V$) in two bins of gas-phase metallicity separated at $12+\log({\rm O/H})=8.5$. We matched the Balmer optical depth ({\taub}) and sSFR distributions of the two samples to control for the dust optical depth biases and different intrinsic stellar populations, respectively.
We investigated the variation of the UV extinction bump at 2175\,{\AA} and the UV slope of the curve in the two metallicity samples. The main results are as follows.

\begin{itemize}
\item The low-metallicity curve appears to show a steep UV rise similar to that of the SMC curve, along with some evidence for variations from galaxy to galaxy. The slope of the high-metallicity curve is identical to that of the Calzetti curve. This result can be explained by different dust grain properties in low- and high-metallicity environments, or different dust-star geometries in young (and low metallicity) galaxies compared to older (more metal rich) galaxies. In the latter scenario, the steep attenuation curve is a result of a higher fraction of young stars embedded in their birth clouds and subject to larger attenuation at short wavelengths.
The former scenario suggests that the small dust grains that produce the UV rise are preferentially present in the low metallicity environments, due to shattering of the large grains in the intense ionizing radiation and/or because of the low fraction of molecular clouds for dust grains to hide in (Sections~\ref{sec:slope} and \ref{sec:imp-dust}).

\item The high-metallicity curve is best fitted with a UV bump at 2175\,{\AA} with $\sim 0.5$ the strength of the MW UV bump. This is a lower limit on the intrinsic strength of the extinction bump, as the bump can be easily suppressed by radiative transfer effects in a clumpy or dusty ISM. A bump of similar strength is ruled out for the low-metallicity curve at $>3\sigma$, showing that the dust grains that produce the bump are significantly less prevalent in the low-metallicity environment. At metallicities of $12+\log({\rm O/H})<8.5$, \citet{shivaei17} showed a significant decrease in the intensity of the mid-IR PAH emission in our sample, which may suggest the PAHs as the main carriers of the bump (Sections~\ref{sec:bump} and \ref{sec:imp-dust}).

\item We propose metallicity/mass dependent attenuation curves to be used for galaxies below and above $12+\log({\rm O/H})\sim 8.5$ and $M_*\sim 10^{10.4}\,M_{\odot}$ (note that our metallicity threshold is based on [N{\sc ii}]/{\halpha} strong line calibrations of \citet{pp04}): at low metallicity/mass, the UV shape of the curve is set to the SMC curve from \citet{gordon03}, and at high metallicity/mass the shape of the curve is set to the \citet{calzetti00} curve with a UV bump added. Adopting the long wavelength shape and $R_V$ of the \citet{reddy15} curve, we propose metallicity-dependent total attenuation curves in Equation~\ref{eq:klam} (Section~\ref{sec:total} and Figure~\ref{fig:totalcurves}).

\item The SFRs inferred from SED model fitting, assuming our two metallicity-dependent attenuation curves, agree well with the independently derived dust-corrected {\sfr} (Section~\ref{sec:sfr}).

\item As the UV bump is a distinct and broad feature that may be significant at high masses, it is advisable to exclude the wavelength region of the bump ($\lambda=1950-2400$\,{\AA}) in calculating the UV continuum slope $\beta$. The effect of a UV bump, similar to the bump we found in the high-metallicity curve, on the future {\em JWST} observations of high-redshift galaxies is presented in Section~\ref{sec:beta}.

\item We find that the ionized gas reddening ($E(B-V)$) is, on average, $\sim 2$ times larger than the stellar continuum reddening at low metallicities, and is similar to the continuum reddening at high metallicities. There is not a strong correlation between the nebular-to-stellar reddening ratio and stellar mass, however the scatter in the reddening ratio increases at low masses, low metallicities, and low SFRs (based on UV diagnostics). These trends may be explained by a physical picture in which at high metallicities, the high surface density of optically thick dust clouds affects both the ionized gas and stellar continuum emission, while the larger scatter in the reddening ratio at lower metallicities, lower masses, and lower SFRs, indicates that the continuum and nebular emission are reddened by different dust components, i.e., the diffuse ISM dust and dusty birthclouds, respectively.
Incorporating other direct tracers of dust attenuation, such as IRX (IR to UV luminosities), in the future, will cast more light on the nature of the difference in the stellar and nebular $E(B-V)$s (Section~\ref{sec:ebmv}).

\end{itemize}

In the future, this study can be improved using larger spectroscopic samples that will allow for finer bins in metallicity. Also, resolved studies of dust and UV emission would shed light on differences in reddening between the lines and continuum (e.g., combining ALMA, UV, and {\halpha} observations).
{\em JWST}/NIRSpec and {\em JWST}/NIRCam observations can extend such studies to higher redshifts. 
{\em JWST}/NIRCam will be an important asset to constrain the UV extinction bump in massive high-redshift galaxies efficiently. Moreover, {\em JWST}/MIRI imaging will provide invaluable data to measure the mid-IR emission at $z\sim 2$ and set robust constraints on the total-to-selective attenuation, $R_V$, which will help to accurately compute dust-corrected SFRs and robust physical parameters.  
\\
\\
\noindent 
I.S. thanks Daniela Calzetti, Stephane Charlot, and Karl Gordon for helpful discussions and feedback on this work. I.S. thanks Joel Leja for his input in the early stages of the project. Support for this work was provided by NASA through the NASA Hubble Fellowship grant \# HST-HF2-51420, awarded by the Space Telescope Science Institute, which is operated by the
Association of Universities for Research in Astronomy, Inc., for NASA, under contract
NAS5-26555.
Funding for the MOSDEF survey was provided by NSF AAG grants AST-1312780, 1312547, 1312764, and 1313171 and archival grant AR-13907, provided by NASA through a grant from the Space Telescope Science Institute.
The data presented herein were obtained at the W. M. Keck Observatory, which is operated as a scientific partnership among the California Institute of Technology, the University of California and the National Aeronautics and Space Administration. The Observatory was made possible by the generous financial support of the W. M. Keck Foundation.
We are grateful to the MOSFIRE instrument team for building this powerful instrument, and to Marc Kassis at the Keck Observatory for his many valuable contributions to the execution of the MOSDEF survey. 
The authors wish to recognize and acknowledge the very significant cultural role and reverence that the summit of Maunakea has always had within the indigenous Hawaiian community. We are most fortunate to have the opportunity to conduct observations from this mountain.

\bibliographystyle{apj}

\appendix

To first order, the behavior of $z\sim 2$ galaxies can be approximated by a smoothly varying rate of star formation \citep[e.g.,][]{reddy12a,shivaei16}. However, stochastic fluctuations of star formation are expected on top of this smooth behavior, as indicated both by observations \citep[e.g.,][]{shivaei15b} and theory \citep[e.g.,][]{sparre17,mitra17}, with amplitudes of 0.2 to 0.4 dex rms in the star-forming SFR-$M_*$ relation. Timescales for fluctuations in the star formation on this scale appear to be of order $\sim 100-200$\,Myr \citep{mitra17,zick18}.

To understand the effects of the star formation history (SFH) on the UV slope, we simulate galaxy stellar populations. Our simulations utilize the PopStar models \citep{molla09} for a Chabrier IMF, which we combine to provide SEDs for different SFHs. These models do not include massive binary stars, which would extend the duration of strong ionizing fluxes by a factor of $\sim 1.5-2$ \citep[e.g.,][]{stanway16}, but would not change our results in a significant way. For the purpose of this section, we measure the UV slope of the composite stellar populations between $1000-2500$\,{\AA}. As a baseline model, we assume a constant star formation of 1\,{\msun}/yr for 1\,Gyr. By definition, this yields a sSFR of $10^{-9}$\,yr$^{-1}$, which is approximately the median sSFR of our sample. The details of the SFH between 200\,Myr and 1\,Gyr have very little effect on the UV slope, and they preserve the sSFR as long as the stellar mass formed prior to 200\,Myr is preserved. 

Over the age range of $0-400$\,Myr, we change the mass of the stellar populations by factors of two over 100\,Myr intervals (including the youngest stellar populations, where these changes result in a change of the sSFR by 0.3\,dex). In addition to the baseline constant SFH model, we also evaluate exponentially rising and declining SFH models with time constants of 200\,Myr, starting at an age of 200\,Myr, and changing the SFR at time zero by factors of two (i.e., changing the sSFR by 0.3\,dex). We find that stars younger than 3\,Myr have a significant effect on the slope, as well as stars between 50 and 200\,Myr. Changes in the SFR in these age ranges by factors of order two could affect the slope by $\sim 10-15$\%. For most other age ranges, the changes are of order $\lesssim 6\%$. For our subsamples of up to 100 galaxies, the changes in UV slope due to stochasticity in the SFR will tend to average out, and the associated net uncertainty in the average slope should be only a few percent.

Another measure of the influence of SED variations on our results can be obtained by ignoring the stochasticity of the SF and adopting the sSFR as a proxy for the shape of the intrinsic SED, assuming a smooth variation of star formation with time. In each metallicity sample, the average sSFR in the {\taub} bins changes by at most $\sim 0.24$\,dex (Figure~\ref{fig:sample-distributions}-right). Assuming a constant star formation history, we estimate the systematic error of the attenuation curve associated with the sSFR difference, as follows. We model two intrinsic (dust-free) SEDs using the \citet{bc03} models, with the mass and SFR of the average mass and SFR in the low and high {\taub} bins in Figure~\ref{fig:sample-distributions}-right. The``correction factor'' that stems from the difference in the two intrinsic SEDs is calculated using Equation~\ref{eq:qi} \citep[based on the error calculation in Section 3.6.4 of][]{reddy15}, and applied to the selective curves ($Q_i$). The ``corrected'' effective curves show $\sim 5$ and  20\% systematic shift to shallower slopes compared to the original high- and low-metallicity curves, respectively, which are within the 1$\sigma$ uncertainties shown in Figure~\ref{fig:att-metal}.

\end{document}